\documentclass[apj]{emulateapj}
\usepackage{times}
\usepackage{natbib}
\usepackage{graphicx}
\usepackage{amsmath}
\usepackage{color}

\newcommand{\str}{Str\"{o}mgren }

\newcommand{\nH}{n_{\rm H, \infty}}

\newcommand{\none}{10~{\rm cm^{-3}}}

\newcommand{\myscale}{1.15}
\newcommand{\msun}{\rm M_\sun}
\newcommand{\mbulge}{M_{\rm bulge}}
\newcommand{\mbulgecrit}{M_{\rm bulge, crit}}
\newcommand{\deltacrit}{\delta_{\rm crit}}
\newcommand{\deltabulge}{\delta_{\rm bulge-BH}}
\newcommand{\mbh}{M_{\rm BH}}
\newcommand{\mbhtwo}{M_{\rm BH}=10^2\,\msun}

\newcommand{\mbhsix}{M_{\rm BH}=10^6\,\msun}
\newcommand{\msigma}{\mbh-\sigma}
\newcommand{\mdotbh}{\dot{M}_{\rm BH}}

\newcommand{\rbeff}{r_{\rm B, eff}}

\shorttitle{Bulge-driven Fueling of Seed Black Holes}
\begin{document}
\title{Bulge-driven Fueling of Seed Black Holes}



\author{KwangHo Park{\altaffilmark{1}}, Massimo Ricotti{\altaffilmark{2}},
Priyamvada Natarajan{\altaffilmark{3}}, Tamara Bogdanovi\'c{\altaffilmark{1}}
, and John H. Wise{\altaffilmark{1}}}
\affil{{\altaffilmark{1}}Center for Relativistic Astrophysics, School of Physics,
Georgia Institute of Technology, Atlanta, GA 30332, USA; \\
kwangho.park@physics.gatech.edu}

\affil{{\altaffilmark{2}}Department of Astronomy, University of Maryland, College
Park, MD 20740, USA}

\affil{{\altaffilmark{3}}Department of Astronomy, Yale University, New Haven, CT 06520, USA} 

\begin{abstract}
We examine radiation-regulated accretion onto intermediate-mass and
massive black holes (BHs) embedded in a bulge component. Using
spherically symmetric one-dimensional radiation-hydrodynamics simulations, we
track the growth of BHs accreting from a cold, neutral gas reservoir
with temperature $T_\infty=10^4$\,K. We find that the accretion rate of BHs
embedded in bulges is proportional to $r_{\rm B,eff}/r_{\rm B}$, where
$r_{\rm B,eff}$ is the increased effective Bondi radius that includes the
gravitational potential of the bulge, and $r_{\rm B}$ is the Bondi
radius of the BH. The radiative feedback from
the BH suppresses the cold accretion rate to $\sim$\,1 percent of
the Bondi rate when a bulge is not considered. However, we find
that the BH fueling rate increases rapidly when the bulge mass
$M_{\rm bulge}$ is greater than the critical value of $\sim 10^6$\,${\rm M}_\sun$
and is proportional to $M_{\rm bulge}$.  Since the critical bulge mass
is independent of the central BH mass, the growth rate of BHs with
masses $M_{\rm BH}=10^2$, $10^4$, and $10^6$\,${\rm M}_\sun$ exhibits distinct
dependencies on the bulge-to-BH mass ratio. Our results imply that
light seed BHs ($\la 10^2$\,${\rm M}_\sun$) which might be the remnants
of the Pop~III stars, cannot grow through accretion coevally with
the early assembly of the bulge of the host galaxies until the bulge
reaches the critical mass. However, massive BH seeds ($\ga
10^5$\,${\rm M}_\sun$) that may form via direct collapse, are more likely
to be embedded in a supercritical bulge and thus can grow efficiently
coupling to the host galaxies and driving the early evolution of
the $M_{\rm BH}-\sigma$ relationship.

\end{abstract}



\section{Introduction}

The discovery of the brightest high redshift quasars in the universe
that are powered by $10^{8}$--$10^9$\,$\msun$ black holes (BHs)
\citep{Fan:2001,Fan:03, Willott:2003, Willott:2010, Fan:2006,
  Mortlock:2011} poses a challenge for models of black hole
formation. How and when the initial seed black holes form and their
subsequent growth history to supermassive BHs (SMBHs) at the centers
of galaxies is an open problem. Recent observations of individual
objects like the 13 billion $\msun$ BH in the quasar detected at
$z=6.3$ \citep{Wu:2015} makes the situation even more challenging to
explain. Given the nature of hierarchical build-up of structure in a
LCDM universe from initial density fluctuations, we can explain a
handful of extreme objects as outliers whose origin can be associated 
with high-$\sigma$ peaks in the density field with peculiar and unusual growth histories
\citep{NatarajanT:2009}. However, the discovery of
populations of bright quasars at progressively earlier epochs in the
universe suggests that a physical explanation is required, one that
might work more generally. With the growing number of discovered
bright quasars at $z > 6$, it is clear that we need a theoretical
model that can produce the progenitor seeds. The traditional
explanation was that the initial BH seeds are the remnants of the
first stars. Such scenarios for the formation of the seed BHs have
suggested that intermediate-mass BHs (IMBHs) with $\mbh \sim
10^2$\,$\msun$ may have formed out of pristine gas as Population~III
star remnants in the early universe
\citep{BrommCL:99,AbelBN:00,MadauR:01} or more massive IMBHs with
$\mbh \sim 10^{4-5}$\,$\msun$ from direct collapse of primordial gas
\citep{Carr:84,HaehneltNR:98,Fryer:01,BegelmanVR:06,JohnsonWFL:2012,
ChoiSB:13,YueFSXC:14,Aykutalp:2014,Inayoshi:2015a} prior to any other structure
formation. Alternatively, gravitationally unstable pre-galactic disks
can give rise to IMBH formation when the gas is only mildly
metal-enhanced \citep{LodatoN:2006, OmukaiSH:08}. Or a primordial star 
cluster may form in the halo and this crowded environment at the core can lead 
to the formation of an IMBH up to $\sim 10^5$\,$\msun$
\citep{Devecchi:2009,Davies:2011,Katz:2015} or to SMBH in high
redshift galaxy mergers \citep{Mayer:2015}.

Thus, the growth rate of IMBHs in the range $10^2$--$10^5$\,~$\msun$ is 
essential to understand the formation and growth history of seed BHs \citep{MadauR:01,
VolonteriHM:03,YooM:04,Volonteri:05,AlvarezWA:09, Natarajan:2011,Jeon:2012}.
Furthermore, radiative feedback from an early population of accreting BHs can significantly 
impact the intergalactic medium \citep{MackOR:07,Ricotti:09}, and the associated luminosities and
spectra provide a potential observational signature of a high-redshift IMBH population.  Studies 
thus far have also provided clues to the origin of ultraluminous X-ray sources
\citep{Krolik:04,RicottiO:04b,StrohmayerM:09}, the BHs at the centers
of dwarf galaxies \citep{Greene:2004,
Reines:2011, Reines:2013, Baldassare:2015}, and potential IMBHs at the centers
of globular clusters \citep[e.g.,][]{Miller:2014,MacLeod:2015}.

Standard accretion theory has difficulty in explaining the rapid
growth from the proposed seed BH masses to the masses powering the
observed bright quasars at high redshift \citep[see][for a
review]{Natarajan:2011,Volonteri:2012}. A brief period of super-Eddington
accretion is often invoked to jump-start BH growth for light seeds.
To compute the accretion rate onto BHs, typically the Bondi accretion
model \citep{BondiH:44,Bondi:52} is applied to estimate the gas
mass accreted from the surrounding medium. Feedback from this
accreting source also needs to be taken into account to estimate
realistic accretion rate, and the resultant luminosity of the BH.
The radiative feedback due to UV and X-ray photons from the BH forms
a hot bubble of gas in the vicinity regulating gas supply from large
scales not only for BHs at rest \citep{ParkR:11, ParkR:12} but also
for BHs in motion relative to the gas \citep{ParkR:13}. These earlier
studies found that the accretion rate drops dramatically to $\sim$ 1
percent of the classical Bondi rate and becomes oscillatory as the
ensuing feedback self-regulates the accretion flow. This is due to
the fact that the gas supply into the \str~sphere is regulated on
the scale of ionization front which is much larger than the Bondi
radius.  The hydrodynamic structure of the \str~sphere changes
constantly due to this self-regulation mechanism.  However, it is
pressure equilibrium across the ionization front that holds the key
to understanding this suppression of the accretion rate. This more
physically realistic treatment, resulting in extremely low accretion
rates, makes it even more challenging to explain the rapid growth
of the seed BH to a few billion $\msun$ at high redshift when the
universe was less than 1\,Gyr old. Our current understanding of the
initial seed BH formation makes it difficult to explain the 4--7
orders of magnitude growth within that short timescale.


Locally we find that the masses of central BHs in galaxies are
tightly correlated to properties of the bulge component. A correlation
between BH mass and bulge velocity dispersion - the $\msigma$
relation appears to hold over many decades in BH mass
\citep{Magorrian:1998,FerrareseM:2000,Tremaine:2002}. Observations
thus reveal that BH properties are closely related to several
properties of their host galaxies. Several models have been proposed
to explain the driving cause for this correlation via energy or
momentum-driven feedback from the central BHs. In these models, the
feedback acts as a thermostat regulating BH growth rate and the
star formation rate of the host galaxy \citep{Silk:98, King:2003,
Murray:2005, SpringelDH:05, OstrikerCCNP:10}.

An alternative explanation may be that the host galaxies drive the
central BH masses by regulating gas supply required for their growth.
In this event, it is important to understand how the BHs respond
to the gas supply since the feedback-regulated accretion is
the key for BH growth.  The associated feedback eventually affects the
properties of the host galaxies as a result. It is clear that we need to 
understand the interplay of the entire growth cycle of BHs and their 
host galaxies.

In this paper, we focus on the role played by the properties of the
bulge in modulating the accretion rate onto the BH. As noted above,
within the idealized Bondi accretion regime, BH growth is highly
suppressed by radiative feedback, therefore it is necessary to
explore other physical scenarios that can potentially boost the
accretion rate. Note that the physical scale of the Bondi radius
of a BH is many orders of magnitude smaller than typical galactic
scales, being $\sim$\,1\,pc for $\mbh=10^4\,\msun$ and $T=10^4$\,K
gas while the average \str~radius is $\sim$\,10\,pc
\citep{ParkR:11,ParkR:12}. This implies that BHs can therefore
only accrete gas from within the compact region - accretion radius
- that is heated by the radiation emitted by the BH. The idealized
Bondi problem setup assumes that the gravitational potential is
determined only by the BH. Here, we examine the role of the other
components that contribute to the gravitational potential near the
BH such as a spheroidal stellar bulge. Seed BHs at the centers of
galaxies do not gravitationally interact with the gas on the larger
galactic scales although the bulge and dark matter halo may play
an important role in delivering gas from large scales to the sub-pc
scales near the BHs.

\begin{figure}[t] \epsscale{\myscale}
\plotone{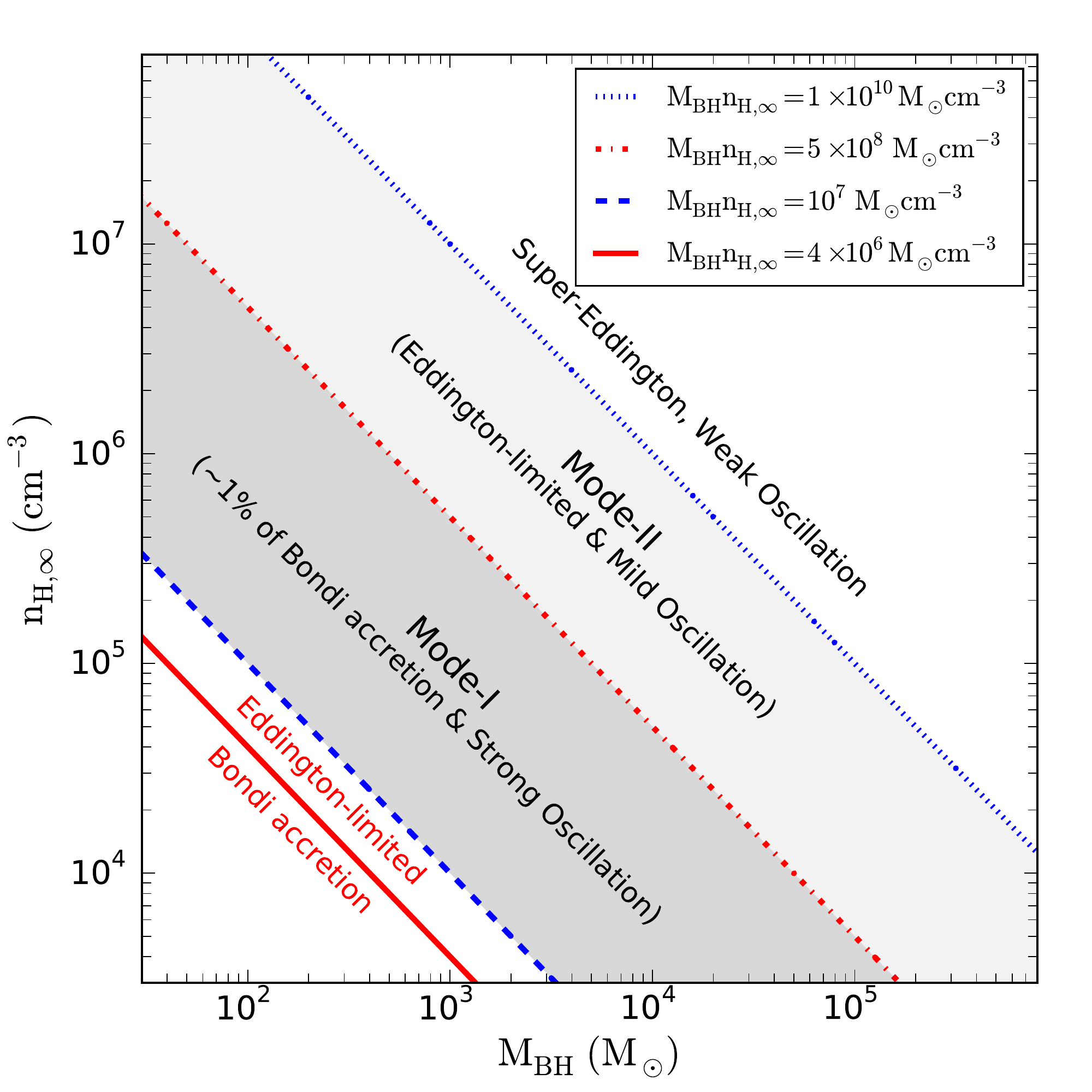} 
\caption{Sketch of BH accretion regimes as a function of BH mass
$\mbh$ and gas density $\nH$. Solid line shows the {\it Eddington-limited
Bondi} recipe where Eddington-limited (above the line) or Bondi
accretion (below) rates are applied for a given $\mbh$ and $\nH$
for $\eta=0.1$ and $T_\infty=10^4$\,K. Two distinct regimes of
Mode-I and Mode-II accretion for a given $\mbh$ and $\nH$ are shown
\citep{ParkR:12}. With increasing $\mbh$ or $\nH$, the accretion
onto the BH reaches the super-Eddington regime. The switch from
Mode-II to super-Eddington regime in the top right corner is an
approximate estimation from \citet{ParkR:12}.} 
\label{fig:mass_den} \end{figure}

\begin{figure*}[t] 
\plotone{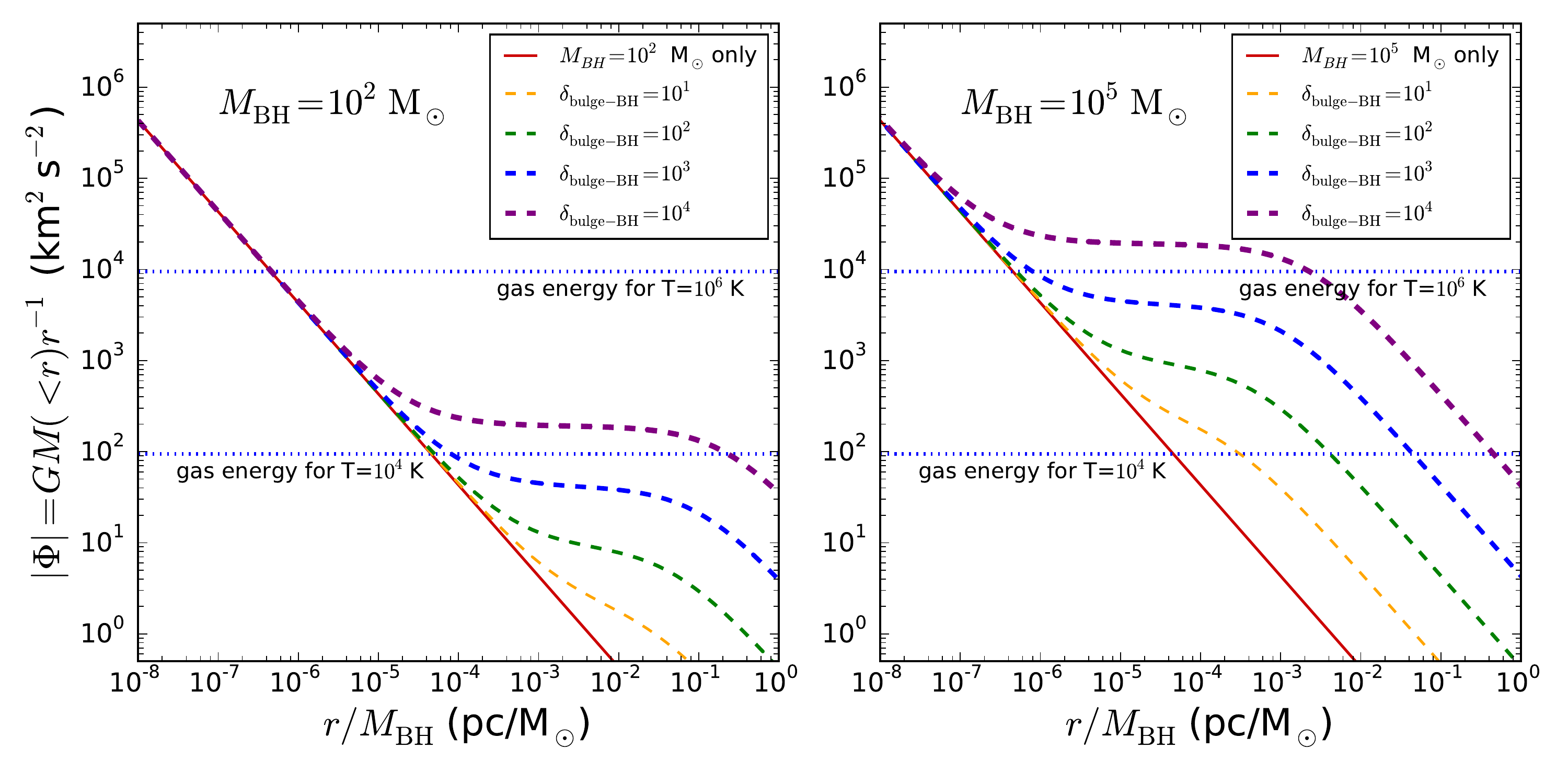} 
\caption{Gravitational potential energy $\Phi$ with a bulge component
as a function of radius for $\mbh=10^2$\,$\msun$ (left) and
$\mbh=10^5$\,$\msun$ (right). The scale radius $a$ for a given
bulge mass is adjusted so that the mass density is fixed within
$a$. The radii are normalized by the BH mass and the gravitational
constant $G=4.3\times 10^{-3}\,$pc\,$\msun^{-1}$(km\,s$^{-1}$)$^2$ is used for $\Phi$. The horizontal dotted lines indicate the
gas energy for $T=10^4$\,K and $10^6$\,K. The intersection of 
$\Phi$ and the gas energy is analogous to {\it Bondi radius} within
which the gravitational energy dominates over the gas energy. Note
that the effective Bondi radius as a function of bulge mass changes
as a function of BH mass and gas temperature.} 
\label{fig:pot_bulge}
\end{figure*}

The main goal of this paper is to explore how the extended mass
distribution around a BH alters the feeding rate of the first BHs.
We investigate the fate of a range of seed
BH masses ($10^2-10^6$\,$\msun$), that are embedded in a neutral
cold ($T_\infty \sim 10^4$\,K) medium. Cold accretion might be more
relevant to the growth of seed BHs in the early universe when the
first galaxies start to build up their stellar mass. We generalize
the results to accretion from a hot-ionized medium ($T_\infty \sim
10^6$\,K) for simulations without radiation feedback. In
section~\ref{sec:method}, we derive new scaling relationships for
the generalized Bondi problem including an extended bulge mass
profile and describe our numerical techniques. In
section~\ref{sec:results}, we present the simulation results and
discuss the results and implications in section~\ref{sec:discussion}.

\section{Methodology}
\label{sec:method}

\subsection{Eddington-limited Bondi accretion}
\label{sec:edd_bondi}
The classical approach adopted to estimate the accretion rate of a
BH is to compute the Bondi rate and apply the Eddington limit while
considering the radiation from the BH. The Bondi radius is defined
as $r_{\rm B} \equiv G\mbh/c_{\rm s, \infty}^2$ where $c_{\rm s,
\infty}$ is the sound speed of the gas. At the Bondi radius, the
gravitational potential energy of gas equals the thermal energy of
the gas, and thus the
gas within $r_{\rm B}$ is accreted to the BH. The Bondi accretion
rate can be expressed as a function of BH mass $\mbh$, ambient
gas density $\rho_\infty$, and sound speed $c_{s, \infty}$ of the gas as
\begin{equation}
\dot{M}_{\rm B} = 4 \pi \lambda_{B} \rho_\infty \frac{G^2 \mbh^2 }{c_{\rm s,\infty}^3}
\end{equation}
where $\lambda_B$ is the dimensionless accretion rate as a function of
the equation of state ($P \propto \rho^\gamma$). The factor
$\lambda_B$ ranges from $e^{3/2}/4$ for an isothermal gas ($\gamma=1$)
to 1/4 for an adiabatic gas ($\gamma=5/3$). The Eddington luminosity
is defined as the maximum accretion rate for a BH with $\mbh$
considering the radiation from the BH and expressed as
\begin{equation}
L_{\rm Edd} = 4\pi G\mbh m_{\rm p}c \sigma_{\rm T}^{-1} 
\label{eq:edd}
\end{equation}
where $m_{\rm p}$ is the proton mass, $\sigma_{\rm T}$ is the Thompson
cross section, and $c$ is the speed of light. The regimes where the
Bondi accretion or Eddington-limited accretion occur can be found by
comparing $\dot{M}_{\rm B}$ and $L_{\rm Edd}/(\eta c^2)$ where $\eta$
is the radiative efficiency. For the case without radiative feedback from the
growing BH,
assuming $\eta=0.1$ and $T_\infty=10^4$\,K the two regimes are
separated by $\mbh\nH=4\times 10^6$\,$\msun {\rm cm}^{-3}$ shown as a
solid line at the bottom left corner of Figure~\ref{fig:mass_den}.

\subsection{Radiation-regulated accretion: Mode-I and II}
\label{sec:two_modes}
When the radiative feedback from the growing BH is taken into
account, the accretion rate is suppressed and displays a highly
oscillatory behavior \citep{MiloCB:09, Li:11, ParkR:11, ParkR:12}.
\citet{ParkR:12} find that two distinct types of oscillations are
expected depending on the values of $\mbh$ and $\nH$, which are
separated by a dot-dashed line shown in Figure~\ref{fig:mass_den}. 
In {\it Mode-I} accretion, the thermal pressure gradient dominates
over the gravity inside the \str~sphere, and bursts of accretion
are driven by the collapse of neutral gas once the gas inside the
\str~sphere is depleted. On the other hand, the oscillation of
accretion rate in {\it Mode-II} is driven by density waves from the
ionization front. The transition between Mode-I (strong) oscillation
with an accretion rate of $\sim$\,1 percent of Bondi rate to Mode-II
(mild) oscillation with Eddington-limited rate occurs approximately at
\begin{equation}
\mbh \nH^{\rm cr} \sim 5 \times 10^8\, \msun\,{\rm cm}^{-3} 
\end{equation}
where $\nH^{\rm cr}$ is the critical density for a given BH mass
$\mbh$. Note that the Eddington-limited accretion regime including
radiation feedback is found at higher $\mbh$ and $\nH$ compared to the
conventional {\it Eddington-limited Bondi accretion} criteria shown in Figure~\ref{fig:mass_den} as a solid red line.
\begin{figure*}[t] 
\plotone{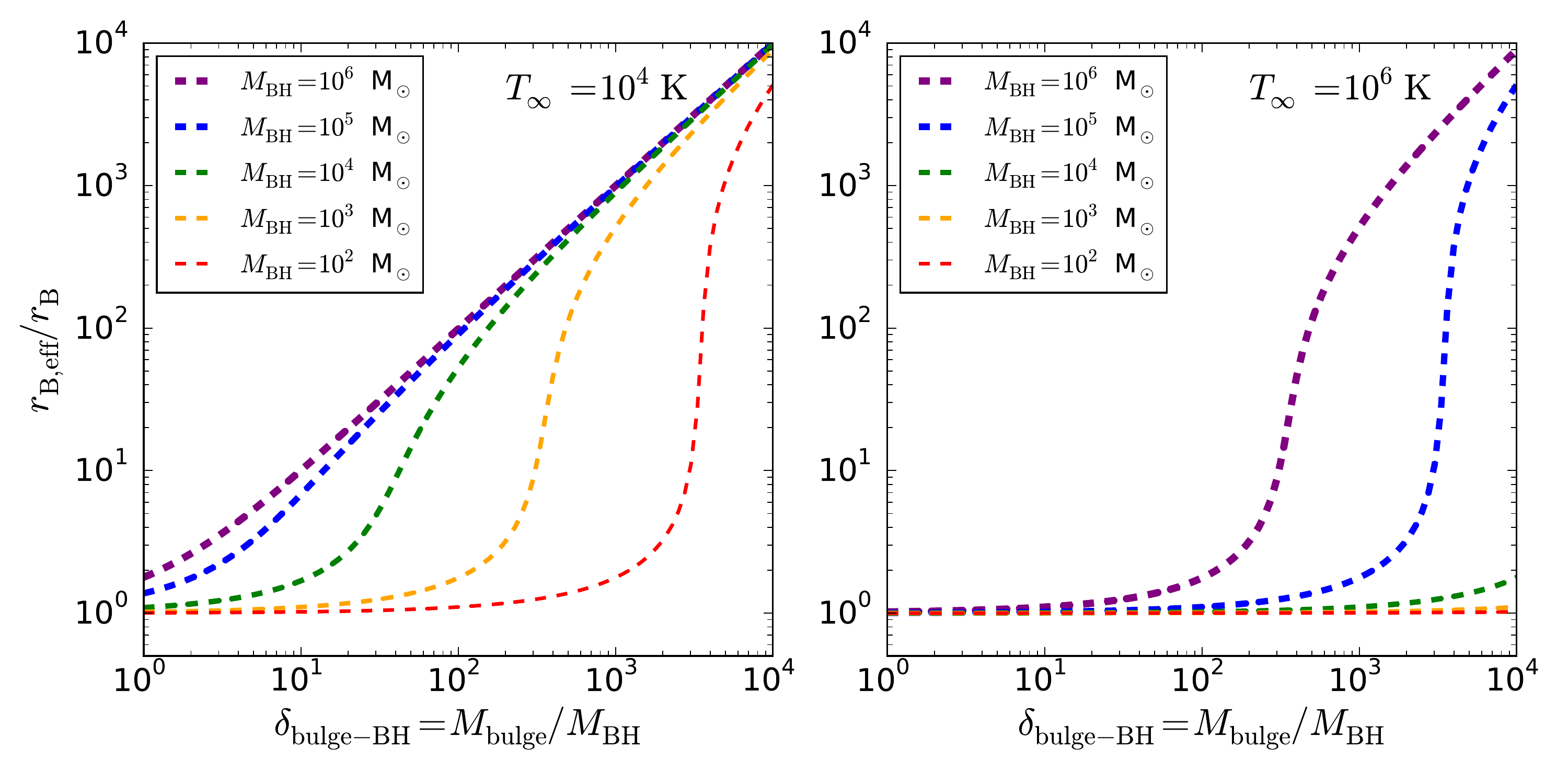} 
\caption{Effective Bondi radius as a function of bulge-to-BH mass
ratio $\deltabulge$ for $\mbh=10^2, 10^3, 10^4, 10^5$, and
$10^6$\,$\msun$. Left panel shows $\rbeff$ for cold gas with
$T_\infty=10^4$\,K and right panel shows $\rbeff$ for hot gas with
$T_\infty=10^6$\,K. } 
\label{fig:r_eff} \end{figure*}

The behavior of the accretion rate is expected to make another transition at 
extremely high densities, large values of $\nH$ for a given $\mbh$, when the 
oscillatory behavior weakens and the radiative feedback is no longer
able to regulate gas accretion. This {\it super-Eddington} or {\it
hyper-accretion} \citep[e.g.,][]{Ohsuga:2011,Jiang:2014} regime
occurs when the
accretion far exceeds the Eddington rate. The dotted line in
Figure~\ref{fig:mass_den} is an upper boundary for
Mode-II accretion covered in \citet{ParkR:12} assuming $\eta=0.1$.
With increasing density for a
given BH mass (moving upward in Figure~\ref{fig:mass_den}) or
increasing BH mass for a fixed gas density (moving to the right in
Figure~\ref{fig:mass_den}), the accretion is expected to make a
transition from the ``{\it feedback-dominated}" to the ``{\it
feeding-dominated}" regime \citep[e.g.,][]{PacucciVF:2015}.
\citet{Inayoshi:2015b} also explore a critical accretion
regime choosing a photon trapping model motivated by \citet{Begelman:79}.
They derive the hyper-accretion regime using the instability of the
ionization front \citep{ParkRDR:14a} when the scale of the photo-ionized region 
becomes smaller the Bondi radius. Of course, physical conditions in the early universe do
permit super-Eddington accretion under certain specific circumstances,
as recently pointed out
by \citet{AlexanderN:2014} when a light BH seed captured in a
dense nuclear star cluster bounces around as it accretes.

\subsection{Generalized Bondi accretion with a bulge component}
\label{sec:r_eff}
We now describe the accretion onto a BH surrounded by a bulge component whose gravitational
potential we now include to derive an analog to Bondi accretion.
To calculate the modified Bondi radius, we model the radial
profile of the stellar distribution with a \citet{Hernquist:90} profile. And the 
corresponding gravitational potential can be described in terms of the total bulge 
mass $\mbulge$ and the scale length $a$ as

\begin{equation}
\Phi_{\rm bulge}(r) = - \frac{GM_{\rm bulge}}{r+a}
\end{equation}
where the enclosed mass within the radius $r$ is $m(r)=M_{\rm
bulge} {r^2} {(r+a)^{-2}}$. The mean density within the scale radius
$a$ is then obtained as 
\begin{equation}
\bar{\rho_*}~(r<a) =\frac{3m(a)}{4\pi a^3} = \frac{3 \mbulge}{16 \pi a^3}.
\end{equation} 


For a given bulge mass, we keep the same density within the scaling
radius $a$ by applying $a= a_0 (\mbulge/\msun)^{1/3}$. 
The Milky Way for instance has a bulge mass $\mbulge \sim 10^{10}\,\msun$
with scale radius $a_{\rm MW} \sim 800$\,pc and the DM halo
mass $ M_{\rm DM} \sim 10^{12}~\msun$
\citep{Dwek:95,Widrow:2005,Kafle:2014}. The average stellar density
within the scale radius for the MW is estimated to be $\sim 1$\,$\msun
{\rm pc}^{-3}$. Here we use $a_0=0.23$\,pc which corresponds to
$\bar{\rho_*} \sim 5.2$\,$\msun {\rm pc}^{-3}$. The $\bar{\rho_*}$
is obviously a free parameter, however note that $a_0$ is not
very sensitive to $\bar{\rho_*}$ since $a_0 \propto \bar{\rho_*}^{-1/3}$
for a given bulge mass.

We define the effective Bondi radius $\rbeff$ as 
\begin{equation}
\frac{G\mbh}{r_{\rm B, eff}}+\frac{G\mbulge}{\rbeff+a} \equiv c^2_\infty 
\label{eq:def_r_eff}
\end{equation}
where the left side of the equation is the magnitude of the
gravitational potential per unit mass due to the BH and bulge
component, whereas the right side represents the thermal energy of the
gas per unit mass.  Figure~\ref{fig:pot_bulge} shows the magnitude of
the combined gravitational potential $\Phi$. The horizontal dotted
lines indicate the specific thermal energy of cold gas with
$T_\infty$=$10^4$\,K and hot gas with $T_\infty=10^6$\,K more
appropriate for massive halos
\citep{McKee:1977,Springel:2003,Pelupessy:07}.  Its intersection with
$\Phi$ for various values of $\deltabulge \equiv \mbulge/\mbh$ is the
solution for the effective Bondi radius $\rbeff$. The solution for the
effective radius in turn can be obtained as

\begin{equation}\begin{split}
&\frac{\rbeff}{r_{\rm B}} =0.5\times \\ \label{eq:r_eff}
&\left[\deltabulge+1-a' + \sqrt{(a'-1-\deltabulge)^2+4a'}\right] 
\end{split}\end{equation}
where the radii are normalized by the Bondi radius as $a' \equiv a/r_{\rm
B}$ and $\deltabulge$ is defined as the bulge-to-BH mass ratio
$\mbulge/\mbh$. 

Figure~\ref{fig:r_eff} shows the effective Bondi radius normalized
by the Bondi radius $\rbeff/r_{\rm B}$ for different BH
masses $10^2$, $10^3$, $10^4$, $10^5$, and $10^6$\,$\msun$. For
cold gas with $T_\infty=10^4$\,K (left panel), the $\rbeff/r_{\rm
B}$ for light BHs with $\mbh=10^2$\,$\msun$ does not increase significantly for
$\deltabulge \la 10^3$ whereas $\rbeff/r_{\rm B}$ for the massive
IMBH with $\mbh \ga 10^5$\,$\msun$ monotonically increases with
$\deltabulge$. This implies that the accretion onto various BH masses
is affected differently with the same $\deltabulge$
since the gas within $\rbeff$ is pulled in to the BH by the enhanced
gravitational potential.


\begin{figure}[t] 
\epsscale{\myscale}
\plotone{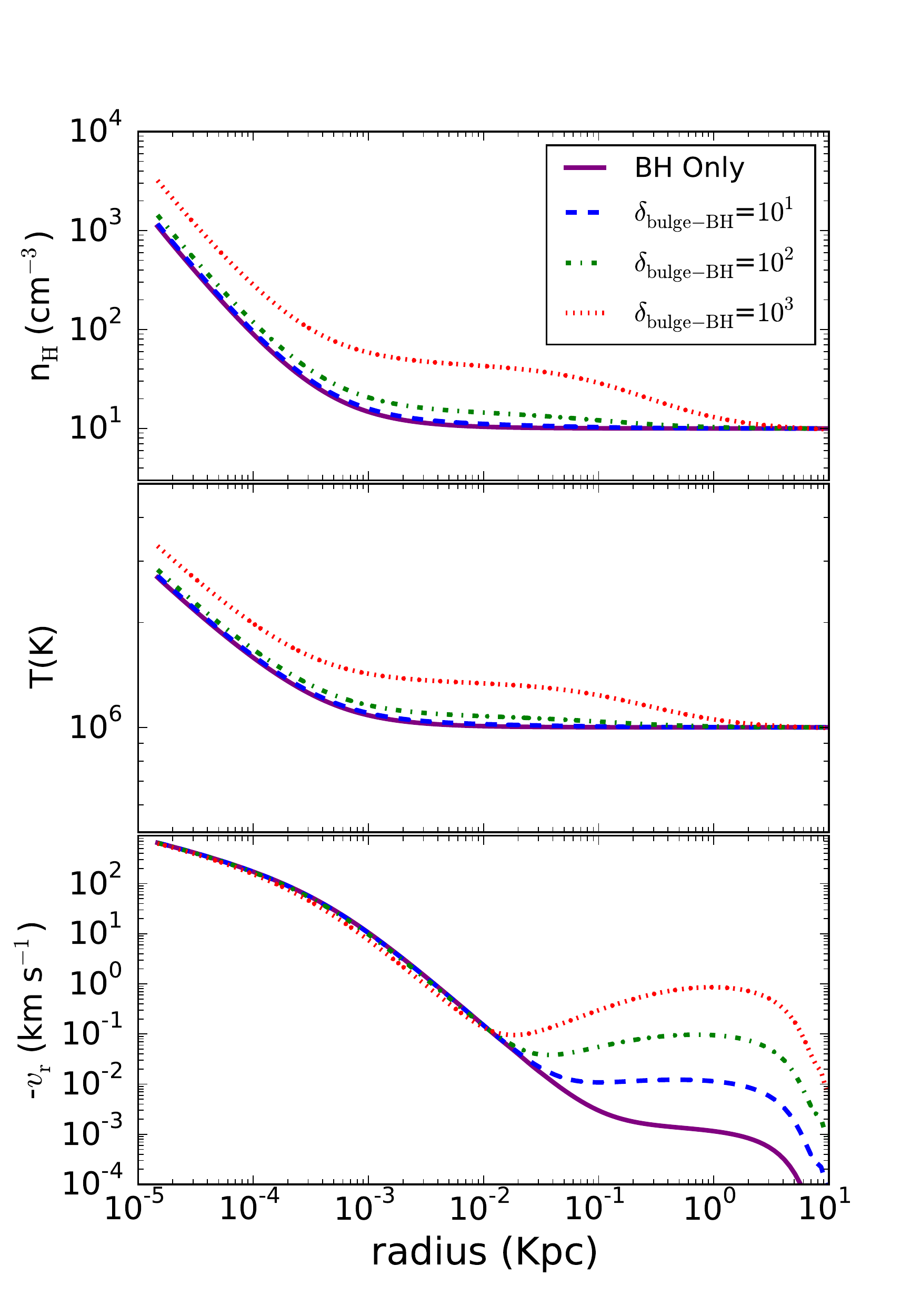} 
\caption{Density (top), temperature (middle), and radial velocity
(bottom) as a function of radius for simulations (M6N1T6NR) without radiative
feedback for $\mbh = 10^6$\,$\msun$, $\nH=\none$, $T_\infty=10^6$\,K,
and $\gamma=1.2$. All the profiles shown are the steady states for
$\deltabulge=0$ (solid), $10^1$
(dashed), $10^2$ (dot-dashed), and $10^3$ (dotted). Inflow velocity
at large radius increases as a function of $\deltabulge$, however
the velocity is at small radius is determined by the gravitational
potential by the BH. The density and temperature profiles do not
change significantly until $\deltabulge \la 10^2$, but shows an enhancement at
$\deltabulge = 10^3$ which is consistent with the behavior of
$\rbeff$ for $T_\infty=10^6$\,K in Figure~\ref{fig:r_eff}.}
\label{fig:profile_norad} \end{figure}

\subsection{Critical Bulge Mass}
From Equation~(\ref{eq:r_eff}), as illustrated in Figure~\ref{fig:r_eff}, it is
evident that there exists a critical bulge mass above which $\rbeff/r_{\rm B}$
transitions from unity to being linearly proportional $\deltabulge$:
thus from the BH Bondi radius to what we now term the ``bulge'' Bondi radius.  

The critical value of the bulge mass can be derived from
Equation~(\ref{eq:r_eff}) noting that $a^\prime \equiv a/r_{\rm B} = C
\deltabulge$, where $C=[a_0/r_{\rm B} (1~\msun)](\mbulge/1~\msun)^{-2/3}$, and 
the transition between $\rbeff/r_{\rm B}$ of unity to $\deltabulge$ happens roughly when $C=1$, i.e., when
\begin{equation}
a=r_{\rm B}\deltabulge=\frac{G\mbulge}{c_{s,\infty}^2}.
\end{equation}
Setting $C=1$, we derive the critical bulge mass at which the
gas accretion rate onto the BH transitions from the case of an
isolated BH to being force-fed by the bulge:
\begin{equation}
\begin{split}
&\mbulgecrit \sim \\ & 4\times 10^5~\msun
\left(\frac{T_\infty}{10^4\,{\rm  K}} \right)^{3/2}
\left(\frac{\bar{\rho_*}}{5.2\,\msun {\rm pc}^{-3}} \right)^{-1/2}.
\label{eq:deltacrit}
\end{split}
\end{equation}

Figure~\ref{fig:r_eff} shows that the value of $\deltacrit \equiv
\mbulgecrit/\mbh$ above which the $\rbeff$ shows a linear relationship
with $\deltabulge$ is $\deltacrit \sim 4000$ for $10^2$\,$\msun$ BHs,
$\deltacrit \sim 400$ for $10^3$\,$\msun$ BHs, and $\deltacrit \sim
40$ for $10^4$\,$\msun$ BHs. This is indeed what was expected if there
is a critical bulge mass $\mbulgecrit \sim 4 \times 10^5$~$\msun$ as
given by Equation~(\ref{eq:deltacrit}) for $T_\infty=10^4$\,K .
For hot gas with $T_\infty=10^6$\,K which is more typical of the hot
virialized gas in massive ellipticals or the hot ionized medium in the
ISM, $\rbeff/r_{\rm B}$ for various BH mass is shown in the right
panel of Figure~\ref{fig:r_eff}. Note that the $\rbeff/r_{\rm B}$ for
a given BH mass for hot gas with $T_{\rm hot}$ matches the case
for lower BH mass $\mbh (T_{\rm cold}/T_{\rm hot})^{3/2}$, as in
Equation~(\ref{eq:deltacrit}).  For example, the value of $\rbeff$ for
$\mbh=10^6$\,$\msun$ and $T_\infty=10^6$\,K matches the one for
$\mbh=10^3$\,$\msun$ and $T_\infty=10^4$\,K.

\begin{table*}[thb]
\begin{center}
\caption{Simulation Parameters}
\begin{tabular}{ccccccc}
\hline 
\hline
   & $\mbh$	&  $\nH$             & $T_\infty$	&		&         	&   \\
ID & $(\msun)$	&  $({\rm cm}^{-3})$ & (K)        	&$\gamma$ 	& Rad Feedback	& $\deltabulge$ \\
\hline
M6N1T4NR & $10^6$  & $10^1$ & $10^4$ & 1.2 	& No & $0$, $10^1$, $10^2$, $10^3$, $10^4$ \\
M6N1T6NR & $10^6$  & $10^1$ & $10^6$ & 1.2, 4/3, 1.4 & No & $0$, $10^1$, $10^2$, $10^3$, $10^4$ \\
M2N5 & $10^2$  & $10^5$ & $10^4$ &  5/3  & Yes & $0$, $3$, $10$, $30$, $10^2$, $3\times 10^2$, $10^3$, $3\times 10^3$, $10^4$ \\
M4N3 & $10^4$  & $10^3$ & $10^4$ &  5/3  & Yes & $0$, $3$, $10$, $30$, $10^2$, $3\times 10^2$, $10^3$, $3\times 10^3$ \\
M6N1 & $10^6$  & $10^1$ & $10^4$ &  5/3  & Yes & $0$, $3$, $10$, $30$, $10^2$, $3\times 10^2$, $10^3$ \\
\hline
\end{tabular}
\label{table:para}
\end{center}
\end{table*}

\begin{figure}[t] 
\epsscale{\myscale}
\plotone{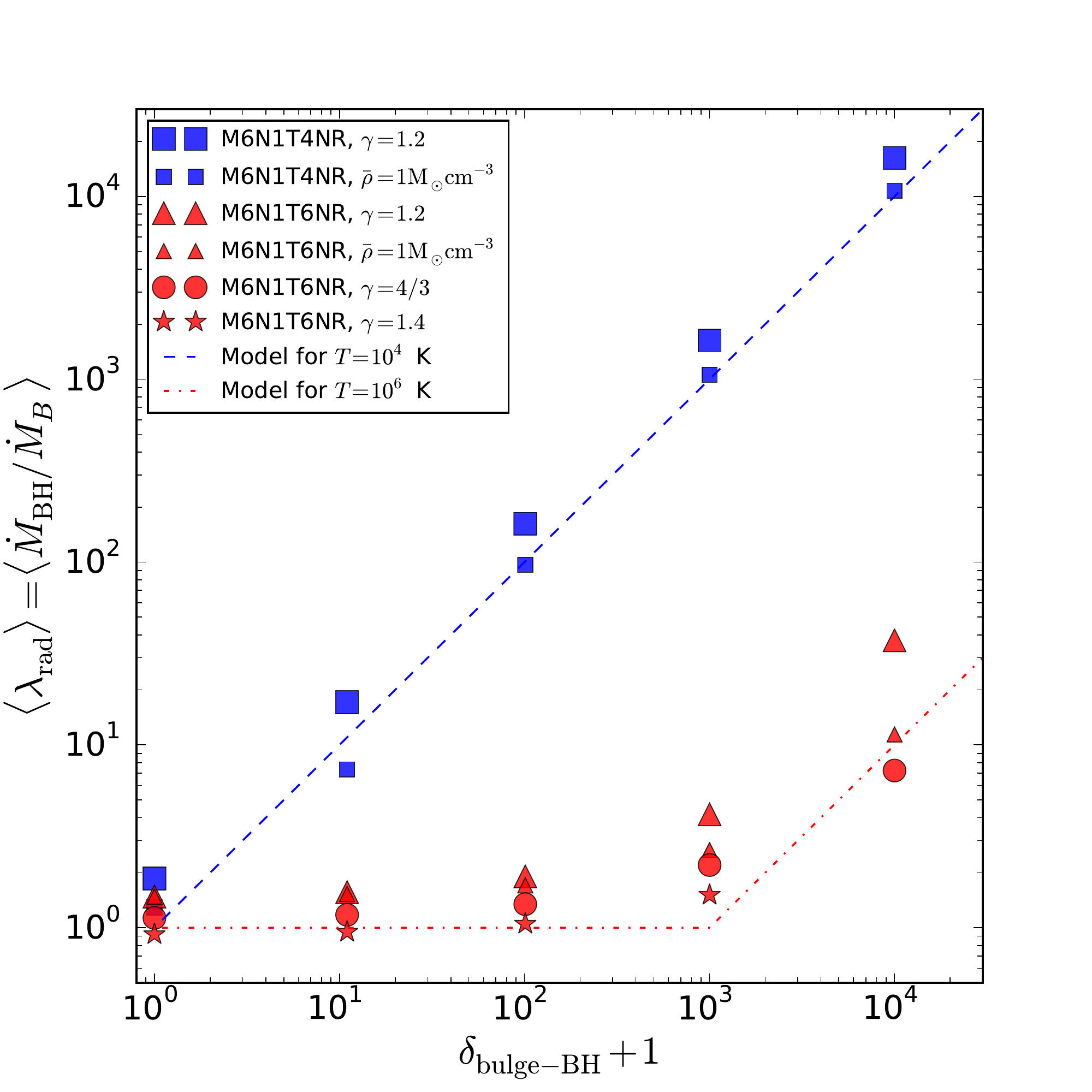} 
\caption{Average accretion rates normalized by Bondi rate for
simulations M6N1T4NR for cold gas ($T_\infty=10^4$\,K) shown as
squares and M6N1T6NR for hot gas ($T_\infty=10^6$\,K) shown as
triangles ($\gamma=1.2$), circles ($\gamma=4/3$), and stars
($\gamma=1.4$). Average accretion rate increases as a function of
$\deltabulge$ when $\deltabulge \ga \deltacrit$. Note that $\deltacrit
\sim 1$ for cold gas while $\deltacrit \sim 10^3$ for hot gas
($T_\infty=10^6$\,K).} 
\label{fig:lambda_norad} \end{figure}

\begin{figure*}[t] 
\epsscale{1.20}
\plotone{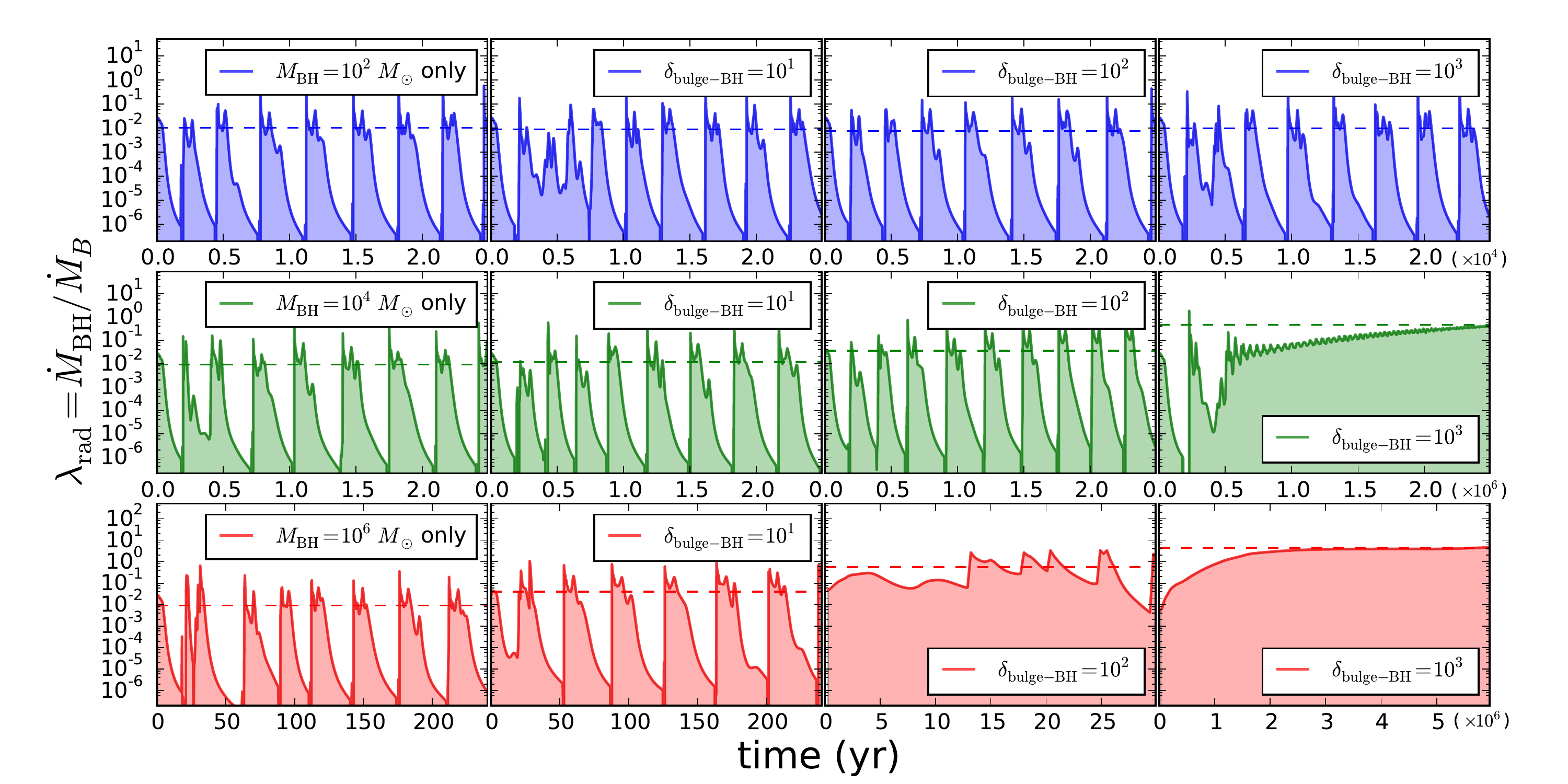} 
\caption{Accretion rates as a function of time for various bulge-to-BH
mass ratios $\deltabulge$. Dashed lines show the mean accretion rates
if the accretion is oscillatory or asymptotic values otherwise. Top
panels show the simulations for M2N5, middle panels show M4N3, and
bottom panels show M6N1. For various $\deltabulge$, M2N5 dose not
show significant change as a function of $\deltabulge$. For M4N3,
due to the increased effective Bondi radius $\rbeff$ the accretion
rate increases as a function of $\deltabulge$ when $\deltabulge \ga
10^2$. For M6N1, the accretion rate increases when $\deltabulge \ga
1$.}
\label{fig:imbh_acc}
\end{figure*}

\subsection{1D Radiation-hydrodynamic Simulations}
\label{sec:sim}
Radiation-hydrodynamic simulations are a useful tool to investigate
the complex interplay between accretion flows and radiative feedback
in the modified Bondi problem with a bulge component.
In this section, we describe the numerical procedures used in our study. 
We run a set of 1D radiation-hydrodynamics simulations using ZEUS-MP
\citep{StoneN:92,Hayes:06} with a radiative transfer equation solver
\citep{RicottiGS:01}. We use a spherical coordinate system with 
a BH centered at $r=0$ applying an operator-splitting
method between hydrodynamic and radiative transfer calculations.
At the minimum radius,
we use the mass flux ($\mdotbh$) to define the BH luminosity
as $L_{\rm bh} = \eta \mdotbh c^2$. 
We apply a power-law spectrum $F_\nu \propto \nu^{-\alpha}$ where
$\alpha$ is the spectral index for BH radiation in the energy range
from $13.6$~eV to $100$~keV and $\alpha=1.5$ is used. Our radiative transfer subroutine
calculates photo-heating, photo-ionization, radiation pressure,
and gas cooling. Compton heating is neglected in this study since the
effect is not significant when the incident spectrum is soft in high accretion
rate regime \citep{ParkRDR:14b}.

The basic setup of the current work is similar to the previous works
\citep{ParkR:11, ParkR:12}, but we add a bulge component to the
gravitational potential (see section~\ref{sec:r_eff}).  Different
pairs of values for $\mbh$ and $\nH$ are selected, but we keep $\mbh
\nH = 10^7$\,$\msun {\rm cm}^{-3}$, so that the we can separate the
effect of the bulge on the growth history of different BH seed masses.
Simulations with the same value of $\mbh \nH$ show qualitatively
similar results in terms of the accretion rate normalized by the Bondi
rate and the period of oscillation when normalized by $\mbh$. Also,
our assumption ensures that the growth timescale of BHs accreting at
the Bondi rate, $\mbh/\dot{M}_{\rm B} \propto \mbh \nH$, is kept
constant in all simulations with different BH masses. Simulation
parameters are listed in Table~\ref{table:para}. M2N5, M4N3, and M6N1
are simulations with radiative feedback for BHs with $10^2$, $10^4$,
and $10^6$\,$\msun$ and keeping the same $\mbh \nH$. M6N1T4NR and
M6N1T6NR are simulations without radiative feedback for
$\mbh=10^6$\,$\msun$, $\none$, and $T_\infty=10^4$\,K and $10^6$\,K,
respectively.

\section{Results}
\label{sec:results}
\subsection{Generalized Bondi Accretion without Radiative Feedback}
Figure~\ref{fig:profile_norad} shows the density (top), temperature
(middle), and radial velocity of the gas (bottom) as a function of
radius for simulations M6N1T6NR without radiative feedback with
$\gamma=1.2$ and $\mbh=10^6$\,$\msun$ accreting from a gas with
temperature $T=10^6$~K. The radial profiles have reached steady state
accretion and the different colored lines (see legend) refer to
different bulge masses: $\deltabulge= 10^1$, $10^2$, and $10^3$ (i.e.,
$\mbulge=10^7, 10^8$, and $10^9$$\msun$).
The density and temperature profiles do not change until $\mbulge
\la 10^8$, but display an enhancement at $\mbulge = 10^9$, that is
consistent with $\mbulgecrit$ in Equation~(\ref{eq:deltacrit}) for
$T_\infty=10^6$\,K and with Figure~\ref{fig:r_eff}.


\begin{figure}[t] 
\epsscale{\myscale}
\plotone{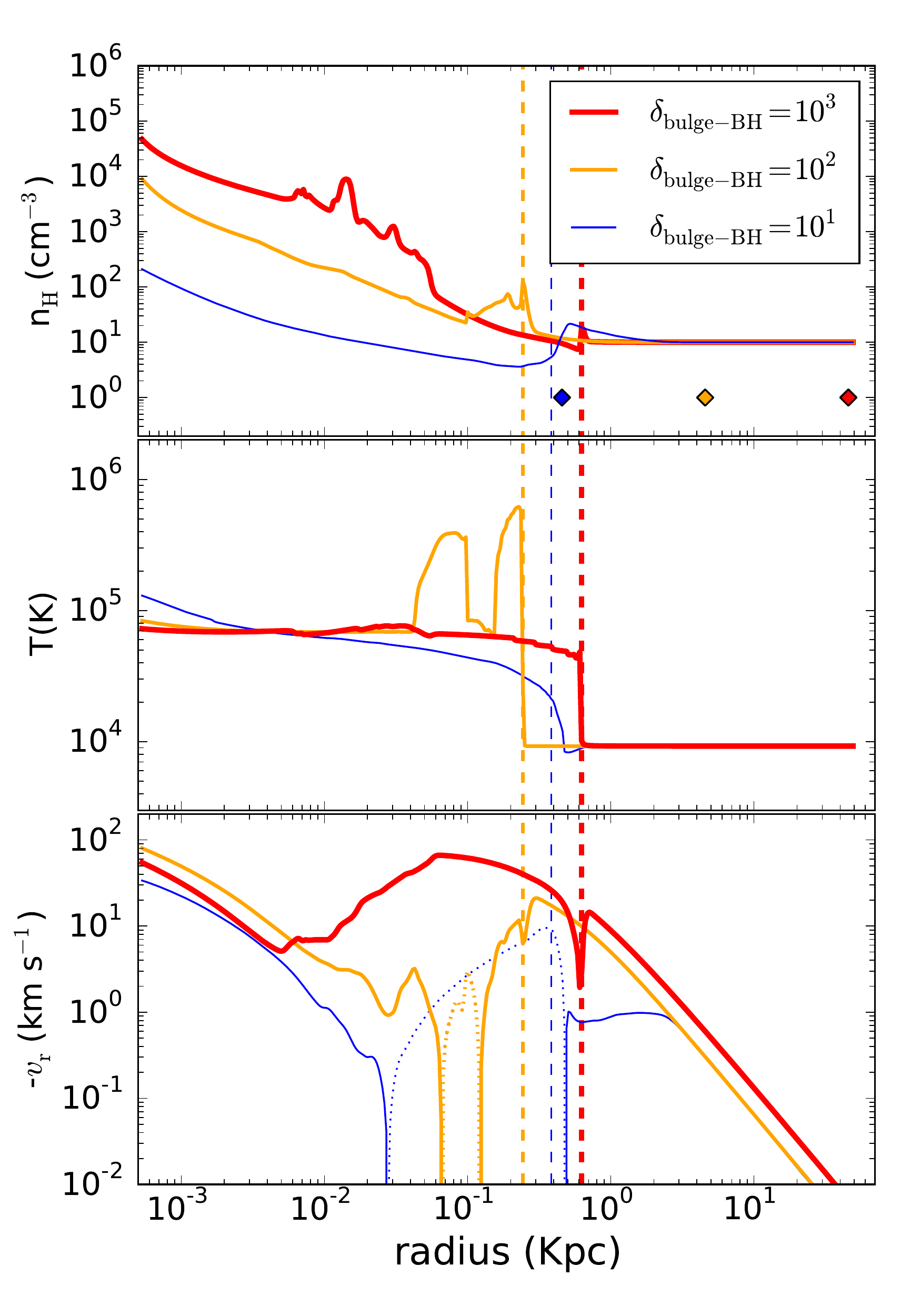} 
\caption{Time-averaged hydrogen number density (top), temperature
(middle), and radial velocity (bottom) profiles as a function of
radius for $\mbh=10^6$\,$\msun$ and $\nH=\none$ simulations.
For $v_r$, solid line is inflow while outflow is shown as dotted. 
Different
colors show the runs with $\deltabulge= 10^1, 10^2$, and
$10^3$, respectively. Vertical dashed lines show the time-averaged
mean size of the \str~sphere for each simulation. Diamonds in the
top panel from left to right indicate $\rbeff$ for simulations with
$\deltabulge=10^1, 10^2$, and $10^3$, respectively. 
Note that $\rbeff$ for $\deltabulge= 10^1$ is comparable to the mean
size of the \str~sphere. As the $\deltabulge$ increases, the
$\rbeff$ becomes larger than the \str~radius.}
\label{fig:profile}
\end{figure}

The change of accretion rate observed when $\mbulge > \mbulgecrit$
for the simulation without radiation feedback appears to be dominated
by an increase in the density produced by the presence of the bulge
component. While the velocity near the BH is not altered as shown in
Figure~\ref{fig:profile_norad}. The central temperature rises
mildly, reducing by the same magnitude of the Bondi radius. Therefore the
dominant effect that causes an increase in the accretion rate is the enhanced
density near the Bondi radius of the BH.

We find that the accretion rate is
$\dot{M}_{\rm B}$ if $\mbulge \le \mbulgecrit$ and increases as
\begin{equation}
\mdotbh \sim  \dot{M}_{\rm B} \frac{\mbulge}{\mbulgecrit},
\label{eq:mdot_bulge}
\end{equation}
for $\mbulge>\mbulgecrit$.
Figure~\ref{fig:lambda_norad} shows accretion rates as a function of
the bulge mass for a set of simulations of $10^6$\,$\msun$ BHs
without radiative feedback but with different values for $\gamma$, the 
temperature of the gas, and density of the bulge (see simulations M6N1T4NR and
M6N1T6NR in Table~\ref{table:para}). We find that
Equation~(\ref{eq:deltacrit}) and (\ref{eq:mdot_bulge}) 
accurately describe the accretion rate from the simulations: the
accretion rate remains constant when $\mbulge \le \mbulgecrit$
($\deltabulge \le \deltacrit$) while it increases linearly with
$\mbulge$ for $\mbulge > \mbulgecrit$ ($\deltabulge >
\deltacrit$). As mentioned in section~\ref{sec:r_eff}, simulations
with a lower stellar density $\bar{\rho_*} \sim 1$\,$\msun {\rm
pc}^{-3}$ (shown as small squares and triangles) do not show a
significant difference from simulations with $\bar{\rho_*} \sim
5.2$\,$\msun {\rm pc}^{-3}$ since the bulge scale length $a$ is not
very sensitive to $\bar{\rho_*}$.

\begin{figure}[t]
\epsscale{\myscale}
\plotone{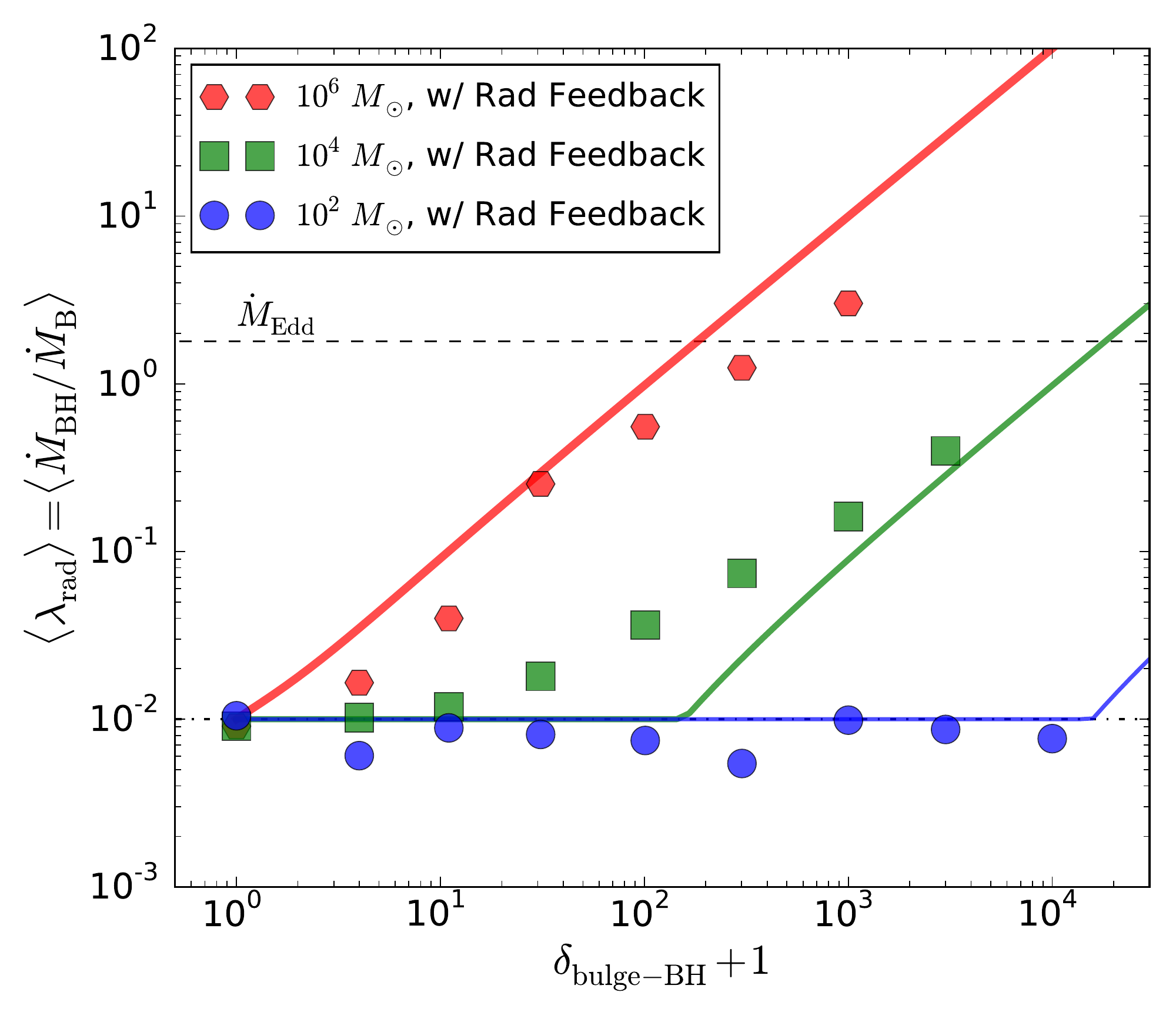} 
\caption{Average accretion rates normalized by Bondi rate for
simulations M2N5 (circles), M4N3 (squares), and M6N1 (hexagons). Average accretion rate increases as a function of
$\deltabulge$ when $\deltabulge \ga \deltacrit$ while accretion
rate remains the same when $\deltabulge \la \deltacrit$ for
all simulations. Since $\deltacrit$ is small ($\sim 1$) for
M6N1, the average accretion rate always increases with increasing bulge
mass $\deltabulge$. On the other hand, the accretion rate does not
increase as a function of of $\deltabulge$ for M2N5 since
$\deltacrit$ is large ($\sim 10^4$).} 
\label{fig:dm_lambda}
\end{figure}

\subsection{Generalized Bondi Accretion with Radiative Feedback}
\label{sec:gen_bondi}
Figure~\ref{fig:imbh_acc} shows accretion rates as a function of time
for $\deltabulge = 0$, $10^1$, $10^2$, and $10^3$ for simulations M2N5
($\mbh=100$~$\msun$, top panels), M4N3 ($\mbh=10^4$\,$\msun$, middle
panels), and M6N1 ($\mbh=10^6$\,$\msun$, bottom panels). The mean
accretion rates are shown as dashed lines in each panel. These 3 sets
of simulations share the same value of $\mbh \nH$ and $T_\infty
=10^4$~K, thus have the same growth timescale and the same
oscillatory behavior in absence of the bulge component (first
columns).  However, including the effect of various bulge masses the
accretion behavior displays significant differences. For a
$10^2$\,$\msun$ BH (M2N5), the assumed $\deltabulge$ produces
negligible effects. The accretion shows oscillatory behavior for the
entire range of $\mbulge$ shown here ($\mbulge \le 10^5$\,$\msun$)
and the mean accretion rate does not change significantly. Therefore,
the presence of a bulge with $\mbulge \le 10^5$\,$\msun$ does not
alter the growth rate of light seed BHs. On the other hand, the
accretion onto a $10^4$\,$\msun$ BH shows a transition from
oscillatory behavior for $\mbulge \la 10^6$\,$\msun$ to quasi-steady
state accretion for $\mbulge \ga 10^7$\,$\msun$. The transition occurs
at a smaller bulge-to-BH ratio $\deltabulge \ga 10^2$ for a
$10^6$\,$\msun$ BH (M6N1), corresponding to $\mbulge \ga
10^8$\,$\msun$. Therefore, to zeroth order, the effect of the bulge on
the mean accretion rate for the case with radiation feedback appear to
be fairly similar to what we found in the case neglecting radiation
feedback. However, the transition from oscillatory behavior to steady
accretion happens for somewhat larger bulge masses than
$\mbulgecrit=10^6$\,$\msun$ when radiative feedback is included and is
mildly dependent on $\deltabulge$. By increasing the bulge mass,
the accretion onto BHs transitions from a duty cycle of $\sim$ 6\%
found for Mode-I \citep{ParkR:12} to a 100\% duty cycle where
the accreting BH is always "on".  
 
Figure~\ref{fig:profile} shows the time-averaged density (top),
temperature (middle), and radial velocity (bottom) profiles as a
function of radius for $\mbh=10^6$\,$\msun$ (run M6N1) with
different bulge masses $\mbulge = 10^7, 10^8$, and $10^9$\,$\msun$
(i.e., $\deltabulge= 10^1, 10^2, $ and $10^3$). The density inside the
\str~radius increases linearly with increasing $\mbulge$ whereas the 
temperature profile only shows a minor change of slope. The
slope of the temperature profile becomes flatter due to the enhanced
density with increasing $\deltabulge$ returning a consistent result
with previous works [e.g., see Figure~8 of \citet{ParkR:11}]. The vertical dashed lines indicate the
average location of the ionization front (i.e.,
\str~radius). The radial velocity profiles also show differences.  In
general, the magnitude of the radial velocity increases as a function
of $\mbulge$. For $\mbulge= 10^7$\,$\msun$,
outflows are clearly seen in the outer part of the \str~sphere (shown as a dotted line), indicating
that the oscillatory accretion rate due to the gas depletion inside
the \str~sphere is still occurring. However, the outflow weakens
with increasing $\mbulge$, leading to a steady accretion
rate. Similar to the case without radiation feedback the increase of
the accretion rate is dominated by the increase in the gas density near
the Bondi radius of the BH. The oscillations instead appear to cease
when $\rbeff$ becomes larger than the \str~radius.

\begin{figure}[t] \epsscale{\myscale}
\plotone{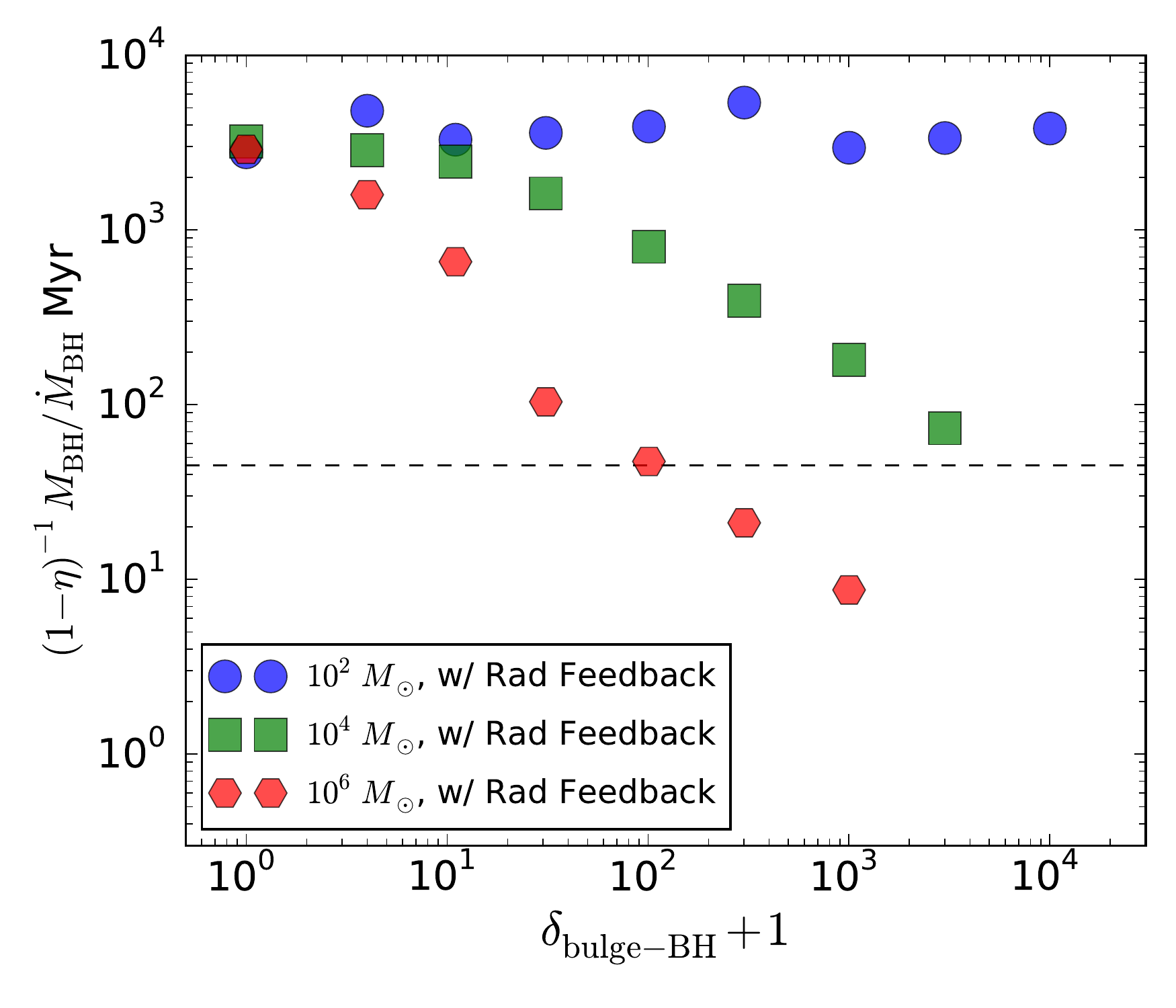}
\caption{Timescales for BH masses $\mbh =10^2$
(circles), $10^4$ (squares), and $10^6$\,$\msun$ (hexagons) to
double the initial mass as a function of $\deltabulge$. 
Salpeter (or {\it e}-folding) timescale
for $\eta=0.1$ accreting at Eddington rate is shown as a dashed
line.} 
\label{fig:slambda} \end{figure}


Figure~\ref{fig:dm_lambda} shows accretion rates normalized by the
Bondi rate for 100\,$\msun$ BHs (runs M2N5, circles),
$10^4$\,$\msun$ BHs (runs M4N3, squares), and $10^6$\,$\msun$ BHs
(runs M6N1, hexagons). The overall accretion rate when radiation
feedback is included is about 1 percent of the Bondi rate, but the
dependence on $\mbulge$ is similar to the case without radiation
feedback (see Figure~\ref{fig:lambda_norad}). Due to the increased
effective accretion radius when $\mbulge \ga \mbulgecrit$, the
accretion rate increases as $\mdotbh \propto \mbulge/\mbulgecrit$
(shown as a solid line). For $10^6$\,$\msun$ BHs (runs M6N1), the
accretion rate increases as approximately $\mbulge$ for $\mbulge \ga
10^6$\,$\msun$, but again at a rate of 1\% compared to the
simulations without radiative feedback.  At $\mbulge \sim
10^7$\,$\msun$ the oscillations are still present, but for $\mbulge
\sim 10^8$\,$\msun$ the accretion rate becomes steady and approaches
$\dot{M}_{\rm B} \sim \dot{M}_{\rm Edd}$ and goes beyond the
Eddington rate at $\mbulge \sim 10^9$\,$\msun$. This suggests a
possibility of the hyper-Eddington accretion regime, however this
work does not establish an upper limit on such high accretion
rates \citep[e.g.,][]{Inayoshi:2015b}. For the runs with smaller
mass BHs (runs M4N3) the same trend of
the accretion rate $\mdotbh \propto
\mbulge/\mbulgecrit$ is found. However, probably due to the
sharper transition of $\rbeff$ from $r_{\rm B}$ to $\deltabulge r_{\rm B}$ as
$\deltabulge$ is increased, the oscillations are damped closer to 
$\mbulge \sim 10^7$\,$\msun$.

Figure~\ref{fig:slambda} shows the timescales for BH masses
with $\mbh=10^2, 10^4$, and $10^6~\msun$ to double the initial mass as
a function of $\deltabulge$. The horizontal line shows the
\citet{Salpeter:64} (or {\it e}-folding) timescale when $\eta=0.1$ and
$L=L_{\rm Edd}$ are assumed. 
For a BH mass $\mbh=10^2$\,$\msun$, the accretion
timescale is $\sim$3\,Gyr with no apparent dependence on
$\deltabulge$. On the other hand, the accretion timescale for
$\mbh=10^6$\,$\msun$ is inversely proportional to the bulge mass
as $\mbh/\mdotbh \propto (\deltabulge + 1)^{-1}$. At $\deltabulge
\sim 10^2$, the accretion timescale for $\mbh=10^6$\,$\msun$ is
similar to the Salpeter timescale. For a BH mass $\mbh=10^4$\,$\msun$,
the accretion timescale does not shows a dependence on $\deltabulge$
for $\deltabulge \la 10^2$ while it decreases with increasing
$\deltabulge$ with the similar slope for $10^6$\,$\msun$.

\section{Discussion and Summary}
\label{sec:discussion}

\begin{figure}[t] \epsscale{\myscale}
\plotone{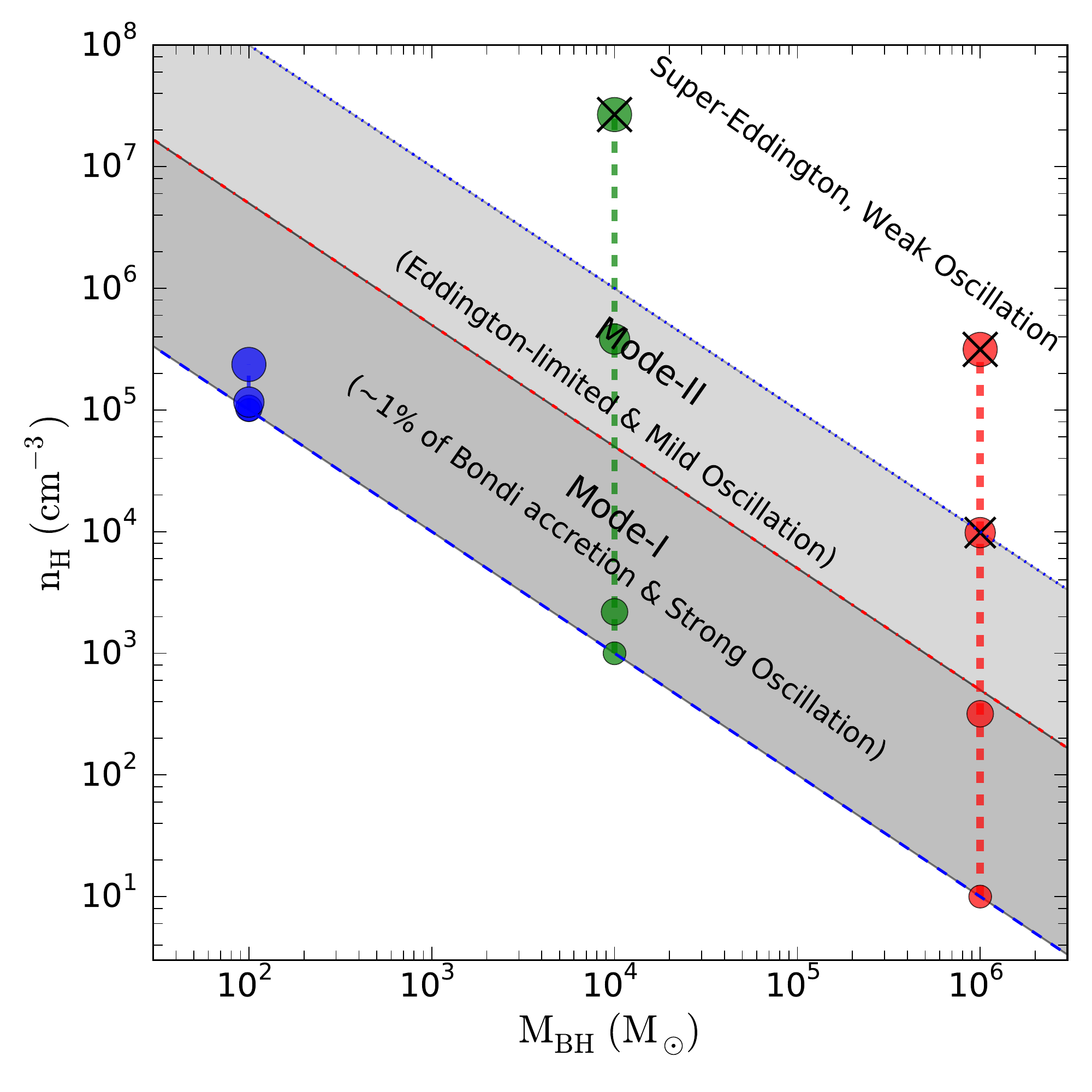} 
\caption{Accretion regimes as a function of $\mbh$ and $\nH$. Symbols
show the shift of accretion regime (from bottom to top) due to the
enhanced density as a function of $\deltabulge$ for BH masses
$10^2$, $10^4$, and $10^6$\,$\msun$. For each BH mass, symbols with
different sizes show $\deltabulge=0, 10^1, 10^2$, and $10^3$ from
bottom to top. Symbols marked with crosses indicate
non-oscillatory cases.}
\label{fig:mn_result} \end{figure}

Accretion regimes as a function of $\mbh$ and $\nH$ explained in
Figure~\ref{fig:mass_den} can be consolidated in presence of the bulge
for various BH masses. In Figure~\ref{fig:mn_result}, the symbols
show the transitions between accretion regimes (from bottom to top)
due to the enhanced density as a function of $\deltabulge$ for BH
masses $10^2$, $10^4$, and $10^6$\,$\msun$. For each BH mass,
symbols with different sizes show $\deltabulge=0, 10^1, 10^2$,
and $10^3$ from bottom to top. Note that for $\mbhsix$, the density
increases linearly with $\deltabulge$, however the density does not
increase for $\mbhtwo$ because $\mbulge<\mbulgecrit \sim 10^6$\,$\msun$
in this case ($T_{\infty}=10^4$~K). Symbols with crosses
in Figure~\ref{fig:mn_result} indicate simulations which do not
show oscillations. 

Although the accretion regime as a function of $\mbh$ and $\nH$ in
Figure~\ref{fig:mn_result} is successful in explaining the transition
of the accretion rates and overall oscillatory behavior presented
in this work, however, the relative location of the \str~radius and
effective Bondi radius also affects the oscillatory behavior
of the accretion. 
For example, a simulation with $\mbh=10^6$\,$\msun$,
$\nH=10^{-2}$\,${\rm cm}^{-3}$, and $\deltabulge=10^3$ where the
effective gas density is enhanced to $\none$ does not show the same
oscillatory behavior as the one with $\mbh=10^6$\,$\msun$,
$\nH=10$\,${\rm cm}^{-3}$ without a bulge. This is due to the same
argument explained in section~\ref{sec:gen_bondi} related to
Figure~\ref{fig:profile} that $\rbeff$ should be smaller than the
average \str~radius for simulations to be oscillatory.

We have focused on accretion from a cold neutral gas reservoir,
which might be the dominant medium in the first galaxies
that form in the early universe. However, accretion onto SMBHs in
evolved galaxies is far more complicated since the gas is multi-phase
with a hot and a cold component, likely
produced by the star formation and the feedback from the SMBH itself.  

This is clearly a first step in understanding the role of the
galactic environment in feeding a seed BH at the center of a galaxy:
we have assumed accretion from a uniform density medium and neglected
angular momentum of the gas that is clearly important for accretion
from the disk in a disk galaxy but is also likely important in
elliptical and spheroidal galaxies \citep{CiottiO:07,CiottiOP:09,NovakOC:11}. In future work,
the growth of seed BHs embedded in various bulge sizes will be studied
in realistic cosmological simulations, guided by findings of 
these idealized simulations.

In this paper, we explore the role of bulge component in
radiation-regulated accretion onto seed IMBHs accreting from cold and
hot medium at the center of primordial galaxies in the early universe.
Here we list our main findings.
\begin{itemize}
\item The presence of a massive bulge makes the effective Bondi
radius $\rbeff$ increase approximately to $\rbeff \simeq r_{\rm B}
\mbulge/\mbh$ for $\mbulge > \mbulgecrit$ while for $\mbulge<
\mbulgecrit$ the bulge has no effect on the accretion radius and
the BH growth.  

\item Without radiative feedback, the accretion rate onto the central
  BH increases as $\mdotbh = \dot{M}_{\rm B} \mbulge/\mbulgecrit$
  when $\mbulge > \mbulgecrit$. This is due to the bulge gravitational
  potential force-feeding the BH and enhancing the density of gas
  near the Bondi radius by a factor $\mbulge/\mbulgecrit$.

\item Including radiative feedback suppresses the accretion rate to 1
  percent of the Bondi rate, however a similar dependence of the
  accretion rate on the bulge mass $\mdotbh \sim 1\% \dot{M}_{\rm
  B} \mbulge/\mbulgecrit$ is found for $\mbulge>\mbulgecrit$. The
  increased gas density near the Bondi
  radius due to fueling from the bulge gravitational potential
  affects the periodic behavior of the accretion rate that eventually
  becomes steady for bulge masses between a few to 10 times
  $\mbulgecrit$.


\item For accretion from a cold gas reservoir, a minimum bulge mass
($\mbulge \sim 10^6$\,$\msun$) exists, above which the effective
Bondi radius is $\rbeff/r_{\rm B} \ga 1$, which is a necessary
condition to boost the BH fueling rate. This might help understanding
the low luminosity for the BHs in small systems such as globular
clusters or dwarf galaxies.  For such systems, the radiative feedback
is effective in suppressing the accretion rate to 1\,percent of the
Bondi rate.

\item For the high mass end of IMBHs, BHs grow fast when $\deltabulge
\ga 1$ meaning that the early development of $\msigma$ relationship
might be a natural consequence of radiation-regulated BH accretion
at the centers of galaxies along with the building of stellar bulge
mass profile. 

\item For the low mass end of IMBHs ($\mbh \sim 10^2$\,$\msun$), BHs
  are harder to grow since the accretion rate is close to 1 percent of
  Bondi rate for $\deltabulge < \deltacrit \sim 10^4$. Thus, assuming
  that the hosts of these small seed BHs have initially $\deltabulge<10^4$
  ($\mbulge<10^6$\,$\msun$ ), the growth timescale for low mass IMBHs
  suggests that light seed BHs, such as Population~III remnants, are
  not able to grow coevally with the bulge.  On the other hand, heavy
  BHs ($\mbh \ga 10^5$\,$\msun$) from direct collapse can grow
  efficiently together with the bulge as long as $\deltabulge \ga
  1- 10$. This result has similar implications about the fate of {\it
    light} and {\it heavy} seed BHs, as the finding of
  \citet{VolonteriN:2009} obtained in models where the $\msigma$
  relation is driven by major mergers.
\end{itemize}

In conclusion, for systems with the bulge masses larger than critical
($\sim 10^6$\,$\msun$ ) the central BH transitions into a new
accretion regime in which both the accretion rate and duty cycle
are maximized despite the presence of radiative feedback. In this
regime, the accretion rate onto the BH reaches and goes beyond the
Eddington rate. While this signals an exciting possibility of
reaching the hyper-Eddington accretion regime, the current work does not
establish an upper limit on such high accretion rates, a topic which
will be explored in future work.

\acknowledgments
This work is supported by the National Science Foundation under the
Theoretical and Computational Astrophysics Network (TCAN) grants
AST-1332858, AST-1333360, AST-1333514. T.B. acknowledges the support
from the Alfred P. Sloan Foundation under Grant No. BR2013-016.

\bibliographystyle{apj}
\bibliography{park_bh}

\begin{thebibliography}{82}
\expandafter\ifx\csname natexlab\endcsname\relax\def\natexlab#1{#1}\fi

\bibitem[{{Abel} {et~al.}(2000){Abel}, {Bryan}, \& {Norman}}]{AbelBN:00}
{Abel}, T., {Bryan}, G.~L., \& {Norman}, M.~L. 2000, \apj, 540, 39

\bibitem[{{Alexander} \& {Natarajan}(2014)}]{AlexanderN:2014}
{Alexander}, T., \& {Natarajan}, P. 2014, Science, 345, 1330

\bibitem[{{Alvarez} {et~al.}(2009){Alvarez}, {Wise}, \& {Abel}}]{AlvarezWA:09}
{Alvarez}, M.~A., {Wise}, J.~H., \& {Abel}, T. 2009, \apjl, 701, L133

\bibitem[{{Aykutalp} {et~al.}(2014){Aykutalp}, {Wise}, {Spaans}, \&
  {Meijerink}}]{Aykutalp:2014}
{Aykutalp}, A., {Wise}, J.~H., {Spaans}, M., \& {Meijerink}, R. 2014, \apj,
  797, 139

\bibitem[{{Baldassare} {et~al.}(2015){Baldassare}, {Reines}, {Gallo}, \&
  {Greene}}]{Baldassare:2015}
{Baldassare}, V.~F., {Reines}, A.~E., {Gallo}, E., \& {Greene}, J.~E. 2015,
  \apjl, 809, L14

\bibitem[{{Begelman}(1979)}]{Begelman:79}
{Begelman}, M.~C. 1979, \mnras, 187, 237

\bibitem[{{Begelman} {et~al.}(2006){Begelman}, {Volonteri}, \&
  {Rees}}]{BegelmanVR:06}
{Begelman}, M.~C., {Volonteri}, M., \& {Rees}, M.~J. 2006, \mnras, 370, 289

\bibitem[{{Bondi}(1952)}]{Bondi:52}
{Bondi}, H. 1952, \mnras, 112, 195

\bibitem[{{Bondi} \& {Hoyle}(1944)}]{BondiH:44}
{Bondi}, H., \& {Hoyle}, F. 1944, \mnras, 104, 273

\bibitem[{{Bromm} {et~al.}(1999){Bromm}, {Coppi}, \& {Larson}}]{BrommCL:99}
{Bromm}, V., {Coppi}, P.~S., \& {Larson}, R.~B. 1999, \apjl, 527, L5

\bibitem[{{Carr} {et~al.}(1984){Carr}, {Bond}, \& {Arnett}}]{Carr:84}
{Carr}, B.~J., {Bond}, J.~R., \& {Arnett}, W.~D. 1984, \apj, 277, 445

\bibitem[{{Choi} {et~al.}(2013){Choi}, {Shlosman}, \& {Begelman}}]{ChoiSB:13}
{Choi}, J.-H., {Shlosman}, I., \& {Begelman}, M.~C. 2013, \apj, 774, 149

\bibitem[{{Ciotti} \& {Ostriker}(2007)}]{CiottiO:07}
{Ciotti}, L., \& {Ostriker}, J.~P. 2007, \apj, 665, 1038

\bibitem[{{Ciotti} {et~al.}(2009){Ciotti}, {Ostriker}, \&
  {Proga}}]{CiottiOP:09}
{Ciotti}, L., {Ostriker}, J.~P., \& {Proga}, D. 2009, \apj, 699, 89

\bibitem[{{Davies} {et~al.}(2011){Davies}, {Miller}, \&
  {Bellovary}}]{Davies:2011}
{Davies}, M.~B., {Miller}, M.~C., \& {Bellovary}, J.~M. 2011, \apjl, 740, L42

\bibitem[{{Devecchi} \& {Volonteri}(2009)}]{Devecchi:2009}
{Devecchi}, B., \& {Volonteri}, M. 2009, \apj, 694, 302

\bibitem[{{Dwek} {et~al.}(1995){Dwek}, {Arendt}, {Hauser}, {Kelsall}, {Lisse},
  {Moseley}, {Silverberg}, {Sodroski}, \& {Weiland}}]{Dwek:95}
{Dwek}, E., {et~al.} 1995, \apj, 445, 716

\bibitem[{{Fan} {et~al.}(2001){Fan}, {Narayanan}, {Lupton}, {Strauss}, {Knapp},
  {Becker}, {White}, {Pentericci}, {Leggett}, {Haiman}, {Gunn}, {Ivezi{\'c}},
  {Schneider}, {Anderson}, {Brinkmann}, {Bahcall}, {Connolly}, {Csabai}, {Doi},
  {Fukugita}, {Geballe}, {Grebel}, {Harbeck}, {Hennessy}, {Lamb}, {Miknaitis},
  {Munn}, {Nichol}, {Okamura}, {Pier}, {Prada}, {Richards}, {Szalay}, \&
  {York}}]{Fan:2001}
{Fan}, X., {et~al.} 2001, \aj, 122, 2833

\bibitem[{{Fan} {et~al.}(2003){Fan}, {Strauss}, {Schneider}, {Becker}, {White},
  {Haiman}, {Gregg}, {Pentericci}, {Grebel}, {Narayanan}, {Loh}, {Richards},
  {Gunn}, {Lupton}, {Knapp}, {Ivezi{\'c}}, {Brandt}, {Collinge}, {Hao},
  {Harbeck}, {Prada}, {Schaye}, {Strateva}, {Zakamska}, {Anderson},
  {Brinkmann}, {Bahcall}, {Lamb}, {Okamura}, {Szalay}, \& {York}}]{Fan:03}
---. 2003, \aj, 125, 1649

\bibitem[{{Fan} {et~al.}(2006){Fan}, {Strauss}, {Becker}, {White}, {Gunn},
  {Knapp}, {Richards}, {Schneider}, {Brinkmann}, \& {Fukugita}}]{Fan:2006}
---. 2006, \aj, 132, 117

\bibitem[{{Ferrarese} \& {Merritt}(2000)}]{FerrareseM:2000}
{Ferrarese}, L., \& {Merritt}, D. 2000, \apjl, 539, L9

\bibitem[{{Fryer} {et~al.}(2001){Fryer}, {Woosley}, \& {Heger}}]{Fryer:01}
{Fryer}, C.~L., {Woosley}, S.~E., \& {Heger}, A. 2001, \apj, 550, 372

\bibitem[{{Greene} \& {Ho}(2004)}]{Greene:2004}
{Greene}, J.~E., \& {Ho}, L.~C. 2004, \apj, 610, 722

\bibitem[{{Haehnelt} {et~al.}(1998){Haehnelt}, {Natarajan}, \&
  {Rees}}]{HaehneltNR:98}
{Haehnelt}, M.~G., {Natarajan}, P., \& {Rees}, M.~J. 1998, \mnras, 300, 817

\bibitem[{{Hayes} {et~al.}(2006){Hayes}, {Norman}, {Fiedler}, {Bordner}, {Li},
  {Clark}, {ud-Doula}, \& {Mac Low}}]{Hayes:06}
{Hayes}, J.~C., {Norman}, M.~L., {Fiedler}, R.~A., {Bordner}, J.~O., {Li},
  P.~S., {Clark}, S.~E., {ud-Doula}, A., \& {Mac Low}, M.-M. 2006, \apjs, 165,
  188

\bibitem[{{Hernquist}(1990)}]{Hernquist:90}
{Hernquist}, L. 1990, \apj, 356, 359

\bibitem[{{Inayoshi} {et~al.}(2015{\natexlab{a}}){Inayoshi}, {Haiman}, \&
  {Ostriker}}]{Inayoshi:2015b}
{Inayoshi}, K., {Haiman}, Z., \& {Ostriker}, J.~P. 2015{\natexlab{a}},
  arXiv:1511.02116

\bibitem[{{Inayoshi} {et~al.}(2015{\natexlab{b}}){Inayoshi}, {Visbal}, \&
  {Kashiyama}}]{Inayoshi:2015a}
{Inayoshi}, K., {Visbal}, E., \& {Kashiyama}, K. 2015{\natexlab{b}}, \mnras,
  453, 1692

\bibitem[{{Jeon} {et~al.}(2012){Jeon}, {Pawlik}, {Greif}, {Glover}, {Bromm},
  {Milosavljevi{\'c}}, \& {Klessen}}]{Jeon:2012}
{Jeon}, M., {Pawlik}, A.~H., {Greif}, T.~H., {Glover}, S.~C.~O., {Bromm}, V.,
  {Milosavljevi{\'c}}, M., \& {Klessen}, R.~S. 2012, \apj, 754, 34

\bibitem[{{Jiang} {et~al.}(2014){Jiang}, {Stone}, \& {Davis}}]{Jiang:2014}
{Jiang}, Y.-F., {Stone}, J.~M., \& {Davis}, S.~W. 2014, \apj, 796, 106

\bibitem[{{Johnson} {et~al.}(2012){Johnson}, {Whalen}, {Fryer}, \&
  {Li}}]{JohnsonWFL:2012}
{Johnson}, J.~L., {Whalen}, D.~J., {Fryer}, C.~L., \& {Li}, H. 2012, \apj, 750,
  66

\bibitem[{{Kafle} {et~al.}(2014){Kafle}, {Sharma}, {Lewis}, \&
  {Bland-Hawthorn}}]{Kafle:2014}
{Kafle}, P.~R., {Sharma}, S., {Lewis}, G.~F., \& {Bland-Hawthorn}, J. 2014,
  \apj, 794, 59

\bibitem[{{Katz} {et~al.}(2015){Katz}, {Sijacki}, \& {Haehnelt}}]{Katz:2015}
{Katz}, H., {Sijacki}, D., \& {Haehnelt}, M.~G. 2015, \mnras, 451, 2352

\bibitem[{{King}(2003)}]{King:2003}
{King}, A. 2003, \apjl, 596, L27

\bibitem[{{Krolik}(2004)}]{Krolik:04}
{Krolik}, J.~H. 2004, \apj, 615, 383

\bibitem[{{Li}(2011)}]{Li:11}
{Li}, Y. 2011, arXiv:1109.3442

\bibitem[{{Lodato} \& {Natarajan}(2006)}]{LodatoN:2006}
{Lodato}, G., \& {Natarajan}, P. 2006, \mnras, 371, 1813

\bibitem[{{Mack} {et~al.}(2007){Mack}, {Ostriker}, \& {Ricotti}}]{MackOR:07}
{Mack}, K.~J., {Ostriker}, J.~P., \& {Ricotti}, M. 2007, \apj, 665, 1277

\bibitem[{{MacLeod} {et~al.}(2015){MacLeod}, {Trenti}, \&
  {Ramirez-Ruiz}}]{MacLeod:2015}
{MacLeod}, M., {Trenti}, M., \& {Ramirez-Ruiz}, E. 2015, arXiv:1508.07000

\bibitem[{{Madau} \& {Rees}(2001)}]{MadauR:01}
{Madau}, P., \& {Rees}, M.~J. 2001, \apjl, 551, L27

\bibitem[{{Magorrian} {et~al.}(1998){Magorrian}, {Tremaine}, {Richstone},
  {Bender}, {Bower}, {Dressler}, {Faber}, {Gebhardt}, {Green}, {Grillmair},
  {Kormendy}, \& {Lauer}}]{Magorrian:1998}
{Magorrian}, J., {et~al.} 1998, \aj, 115, 2285

\bibitem[{{Mayer} {et~al.}(2015){Mayer}, {Fiacconi}, {Bonoli}, {Quinn}, {Ro{\v
  s}kar}, {Shen}, \& {Wadsley}}]{Mayer:2015}
{Mayer}, L., {Fiacconi}, D., {Bonoli}, S., {Quinn}, T., {Ro{\v s}kar}, R.,
  {Shen}, S., \& {Wadsley}, J. 2015, \apj, 810, 51

\bibitem[{{McKee} \& {Ostriker}(1977)}]{McKee:1977}
{McKee}, C.~F., \& {Ostriker}, J.~P. 1977, \apj, 218, 148

\bibitem[{{Miller} {et~al.}(2014){Miller}, {Farrell}, \&
  {Maccarone}}]{Miller:2014}
{Miller}, M.~C., {Farrell}, S.~A., \& {Maccarone}, T.~J. 2014, \apj, 788, 116

\bibitem[{{Milosavljevi{\'c}} {et~al.}(2009){Milosavljevi{\'c}}, {Couch}, \&
  {Bromm}}]{MiloCB:09}
{Milosavljevi{\'c}}, M., {Couch}, S.~M., \& {Bromm}, V. 2009, \apjl, 696, L146

\bibitem[{{Mortlock} {et~al.}(2011){Mortlock}, {Warren}, {Venemans}, {Patel},
  {Hewett}, {McMahon}, {Simpson}, {Theuns}, {Gonz{\'a}les-Solares}, {Adamson},
  {Dye}, {Hambly}, {Hirst}, {Irwin}, {Kuiper}, {Lawrence}, \&
  {R{\"o}ttgering}}]{Mortlock:2011}
{Mortlock}, D.~J., {et~al.} 2011, \nat, 474, 616

\bibitem[{{Murray} {et~al.}(2005){Murray}, {Quataert}, \&
  {Thompson}}]{Murray:2005}
{Murray}, N., {Quataert}, E., \& {Thompson}, T.~A. 2005, \apj, 618, 569

\bibitem[{{Natarajan}(2011)}]{Natarajan:2011}
{Natarajan}, P. 2011, Bulletin of the Astronomical Society of India, 39, 145

\bibitem[{{Natarajan} \& {Treister}(2009)}]{NatarajanT:2009}
{Natarajan}, P., \& {Treister}, E. 2009, \mnras, 393, 838

\bibitem[{{Novak} {et~al.}(2011){Novak}, {Ostriker}, \& {Ciotti}}]{NovakOC:11}
{Novak}, G.~S., {Ostriker}, J.~P., \& {Ciotti}, L. 2011, \apj, 737, 26

\bibitem[{{Ohsuga} \& {Mineshige}(2011)}]{Ohsuga:2011}
{Ohsuga}, K., \& {Mineshige}, S. 2011, \apj, 736, 2

\bibitem[{{Omukai} {et~al.}(2008){Omukai}, {Schneider}, \&
  {Haiman}}]{OmukaiSH:08}
{Omukai}, K., {Schneider}, R., \& {Haiman}, Z. 2008, \apj, 686, 801

\bibitem[{{Ostriker} {et~al.}(2010){Ostriker}, {Choi}, {Ciotti}, {Novak}, \&
  {Proga}}]{OstrikerCCNP:10}
{Ostriker}, J.~P., {Choi}, E., {Ciotti}, L., {Novak}, G.~S., \& {Proga}, D.
  2010, \apj, 722, 642

\bibitem[{{Pacucci} {et~al.}(2015){Pacucci}, {Volonteri}, \&
  {Ferrara}}]{PacucciVF:2015}
{Pacucci}, F., {Volonteri}, M., \& {Ferrara}, A. 2015, \mnras, 452, 1922

\bibitem[{{Park} \& {Ricotti}(2011)}]{ParkR:11}
{Park}, K., \& {Ricotti}, M. 2011, \apj, 739, 2

\bibitem[{{Park} \& {Ricotti}(2012)}]{ParkR:12}
---. 2012, \apj, 747, 9

\bibitem[{{Park} \& {Ricotti}(2013)}]{ParkR:13}
---. 2013, \apj, 767, 163

\bibitem[{{Park} {et~al.}(2014{\natexlab{a}}){Park}, {Ricotti}, {Di Matteo}, \&
  {Reynolds}}]{ParkRDR:14a}
{Park}, K., {Ricotti}, M., {Di Matteo}, T., \& {Reynolds}, C.~S.
  2014{\natexlab{a}}, \mnras, 437, 2856

\bibitem[{{Park} {et~al.}(2014{\natexlab{b}}){Park}, {Ricotti}, {Di Matteo}, \&
  {Reynolds}}]{ParkRDR:14b}
---. 2014{\natexlab{b}}, \mnras, 445, 2325

\bibitem[{{Pelupessy} {et~al.}(2007){Pelupessy}, {Di Matteo}, \&
  {Ciardi}}]{Pelupessy:07}
{Pelupessy}, F.~I., {Di Matteo}, T., \& {Ciardi}, B. 2007, \apj, 665, 107

\bibitem[{{Reines} {et~al.}(2013){Reines}, {Greene}, \& {Geha}}]{Reines:2013}
{Reines}, A.~E., {Greene}, J.~E., \& {Geha}, M. 2013, \apj, 775, 116

\bibitem[{{Reines} {et~al.}(2011){Reines}, {Sivakoff}, {Johnson}, \&
  {Brogan}}]{Reines:2011}
{Reines}, A.~E., {Sivakoff}, G.~R., {Johnson}, K.~E., \& {Brogan}, C.~L. 2011,
  \nat, 470, 66

\bibitem[{{Ricotti}(2009)}]{Ricotti:09}
{Ricotti}, M. 2009, \mnras, 392, L45

\bibitem[{{Ricotti} {et~al.}(2001){Ricotti}, {Gnedin}, \&
  {Shull}}]{RicottiGS:01}
{Ricotti}, M., {Gnedin}, N.~Y., \& {Shull}, J.~M. 2001, \apj, 560, 580

\bibitem[{{Ricotti} \& {Ostriker}(2004)}]{RicottiO:04b}
{Ricotti}, M., \& {Ostriker}, J.~P. 2004, \mnras, 352, 547

\bibitem[{{Salpeter}(1964)}]{Salpeter:64}
{Salpeter}, E.~E. 1964, \apj, 140, 796

\bibitem[{{Silk} \& {Rees}(1998)}]{Silk:98}
{Silk}, J., \& {Rees}, M.~J. 1998, \aap, 331, L1

\bibitem[{{Springel} {et~al.}(2005){Springel}, {Di Matteo}, \&
  {Hernquist}}]{SpringelDH:05}
{Springel}, V., {Di Matteo}, T., \& {Hernquist}, L. 2005, \mnras, 361, 776

\bibitem[{{Springel} \& {Hernquist}(2003)}]{Springel:2003}
{Springel}, V., \& {Hernquist}, L. 2003, \mnras, 339, 289

\bibitem[{{Stone} \& {Norman}(1992)}]{StoneN:92}
{Stone}, J.~M., \& {Norman}, M.~L. 1992, \apjs, 80, 753

\bibitem[{{Strohmayer} \& {Mushotzky}(2009)}]{StrohmayerM:09}
{Strohmayer}, T.~E., \& {Mushotzky}, R.~F. 2009, \apj, 703, 1386

\bibitem[{{Tremaine} {et~al.}(2002){Tremaine}, {Gebhardt}, {Bender}, {Bower},
  {Dressler}, {Faber}, {Filippenko}, {Green}, {Grillmair}, {Ho}, {Kormendy},
  {Lauer}, {Magorrian}, {Pinkney}, \& {Richstone}}]{Tremaine:2002}
{Tremaine}, S., {et~al.} 2002, \apj, 574, 740

\bibitem[{{Volonteri}(2012)}]{Volonteri:2012}
{Volonteri}, M. 2012, Science, 337, 544

\bibitem[{{Volonteri} {et~al.}(2003){Volonteri}, {Haardt}, \&
  {Madau}}]{VolonteriHM:03}
{Volonteri}, M., {Haardt}, F., \& {Madau}, P. 2003, \apj, 582, 559

\bibitem[{{Volonteri} \& {Natarajan}(2009)}]{VolonteriN:2009}
{Volonteri}, M., \& {Natarajan}, P. 2009, \mnras, 400, 1911

\bibitem[{{Volonteri} \& {Rees}(2005)}]{Volonteri:05}
{Volonteri}, M., \& {Rees}, M.~J. 2005, \apj, 633, 624

\bibitem[{{Widrow} \& {Dubinski}(2005)}]{Widrow:2005}
{Widrow}, L.~M., \& {Dubinski}, J. 2005, \apj, 631, 838

\bibitem[{{Willott} {et~al.}(2003){Willott}, {McLure}, \&
  {Jarvis}}]{Willott:2003}
{Willott}, C.~J., {McLure}, R.~J., \& {Jarvis}, M.~J. 2003, \apjl, 587, L15

\bibitem[{{Willott} {et~al.}(2010){Willott}, {Delorme}, {Reyl{\'e}}, {Albert},
  {Bergeron}, {Crampton}, {Delfosse}, {Forveille}, {Hutchings}, {McLure},
  {Omont}, \& {Schade}}]{Willott:2010}
{Willott}, C.~J., {et~al.} 2010, \aj, 139, 906

\bibitem[{{Wu} {et~al.}(2015){Wu}, {Wang}, {Fan}, {Yi}, {Zuo}, {Bian}, {Jiang},
  {McGreer}, {Wang}, {Yang}, {Yang}, {Thompson}, \& {Beletsky}}]{Wu:2015}
{Wu}, X.-B., {et~al.} 2015, \nat, 518, 512

\bibitem[{{Yoo} \& {Miralda-Escud{\'e}}(2004)}]{YooM:04}
{Yoo}, J., \& {Miralda-Escud{\'e}}, J. 2004, \apjl, 614, L25

\bibitem[{{Yue} {et~al.}(2014){Yue}, {Ferrara}, {Salvaterra}, {Xu}, \&
  {Chen}}]{YueFSXC:14}
{Yue}, B., {Ferrara}, A., {Salvaterra}, R., {Xu}, Y., \& {Chen}, X. 2014,
  \mnras, 440, 1263

\end{thebibliography}

\label{lastpage}
\end{document}